\documentclass[usenatbib] {mn2e}
\usepackage{graphicx}
\usepackage{subfig}
\usepackage{float}
\usepackage{times}
\usepackage{natbib}

\newcommand{\enzo}{{ENZO\ }}

\begin{document}

\title[Properties of Filaments]{ 
Evolution of cosmic filaments and of their galaxy population from MHD cosmological simulations}
\author[C. Gheller, F. Vazza, M. Br\"{u}ggen, M. Alpaslan, B. W. Holwerda, A. M. Hopkins, J. Liske]{C. Gheller$^{1}$, F. Vazza$^{2,3}$, M. Br\"{u}ggen$^{2}$, M. Alpaslan$^{4}$, B.W. Holwerda$^{5}$, A.M. Hopkins$^{6}$, J. Liske$^{2}$\\
$^{1}$ ETHZ-CSCS, Via Trevano 131, Lugano, Switzerland \\
$^{2}$ Hamburger Sternwarte, Universit\"at Hamburg, Gojenbergsweg 112, 21029 Hamburg, Germany\\
$^{3}$ INAF-Istituto di Radio Astronomia, Via Gobetti 101, Bologna, Italy\\
$^{4}$ NASA Ames Research Center, N232, Moffett Field, Mountain View, CA 94035, United States\\
$^{5}$ University of Leiden, Sterrenwacht Leiden, Niels Bohrweg 2, NL-2333 CA Leiden, The Netherlands\\
$^{6}$ Australian Astronomical Observatory P.O. Box 915, North Ryde NSW 1670, Australia
 }

\date{Accepted ???. Received ???; in original form ???}
\maketitle

\begin{abstract}
Despite containing about a half of the total matter in the Universe, at most wavelengths the filamentary structure of the cosmic web is difficult to observe. In this work, we use large unigrid cosmological simulations to investigate how the geometrical, thermodynamical and 
magnetic properties of cosmological filaments vary with mass and redshift (z $\leq 1$). We find that the average temperature, length, volume and magnetic field of filaments are tightly log-log correlated with the underlying total gravitational mass.
This reflects the role of self-gravity 
in shaping their properties and enables statistical predictions of their observational properties based on their mass. 
We also focus on the properties of the simulated population of galaxy-sized halos within filaments, and compare their properties to the results
obtained from the spectroscopic GAMA survey. Simulated and observed filaments with the same length are found to contain an equal number of galaxies, with very similar distribution of halo masses. The  total number of galaxies within each filament and the total/average
stellar mass in galaxies can now be used to predict also the large-scale properties of the gas in the host filaments across tens or hundreds of Mpc in scale. 
These results are the first steps  towards the future use of galaxy catalogues in order to select the best targets for observations of the warm-hot intergalactic medium.

\end{abstract}


\label{firstpage} 
\begin{keywords}
galaxy: intergalactic medium, methods: numerical, cosmology: large-scale structure of Universe
\end{keywords}

\vskip 0.4cm \fontsize{11pt}{11pt} \selectfont  

\section{Introduction}
\label{sec:intro}

The large-scale structure of the 
Universe is organized into a web of filamentary matter connecting clusters and 
groups of galaxies and wall--like structures separating gigantic underdense regions 
(voids). A large fraction of the baryonic matter (around 50\%) is predicted to reside in such {\it cosmic web}, 
at densities $\sim 10-100$ times the average cosmic value and temperature of
$10^5-10^7$K, forming the ``Warm-Hot Intergalactic Medium'' (WHIM, see e.g. \citealt[][]{1999ApJ...514....1C,2001ApJ...552..473D},
\citealt[][]{1996Natur.380..603B}, \citealt[and references therein]{do08, 2001ApJ...552..473D,2005MNRAS.360.1110V,2006MNRAS.370..656D}).

Direct observations of the cosmic web are challenging, due to its extremely low mass density.
Indications of possible detections of filamentary structures emerged from the analysis of 
soft X--ray \citep[e.g.][]{2003A&A...410..777F,2008A&A...482L..29W,2010ApJ...715..854N,2013ApJ...769...90N}, 
or Sunyaev-Zeldovich effect \citep[][]{2013A&A...550A.134P} data. However, very recently
 the first X-ray imaging of the terminal part of four filaments connected to the virial radius of cluster A2744 has been
reported by \citet{2015Natur.528..105E} using XMM-Newton.  

The interest in the observational detection of filaments is growing rapidly since
their evolution is less violent and complex than that of galaxy clusters, with adiabatic physics (besides gravity)
dominating the gas dynamics. Therefore, filaments are expected to preserve many
traces of the original environment in which the process of gravitational clustering started. They are also expected to retain
memory of the initial magnetic seed fields of the Universe because they should not host strong dynamo amplification \citep[][]{ry08,donn09,va14mhd}.
Cosmological simulations can now be used to produce observable predictions for different models of extragalactic magnetic fields, to be tested by radio observations. 
Radio surveys at low frequencies (e.g. LOFAR, MWA and SKA)  are expected to detect the brightest parts of the filamentary cosmic web  if the magnetic field in the WHIM is amplified to the level of a few  $\%$ of the thermal gas energy there \citep[][]{va15ska,va15radio}. 
Predicting the evolution of the magnetic fields in filaments is important because radio surveys might be able to detect the cosmic web in the redshift range $z \sim 0.1-0.2$, while especially at high radio frequencies it is impossible to sample the large angular scales ($\geq $ degrees) associated with the emission from the cosmic web in the local Universe \citep[][]{va15radio}.\\

In addition to galaxy-galaxy mergers \citep[e.g.][]{2003AJ....126.1183C,2006ApJ...645..986R,2007A&A...468...61D,2009ApJ...699L.178N} and the co-evolution of supermassive black holes \citep[e.g.][]{2005MNRAS.361..776S,2007MNRAS.382.1415S,2008ApJS..175..356H},
the relation between galaxies and their surrounding large-scale environment is key to understand the accretion of cold gas into galactic centres \citep[e.g.][]{2015MNRAS.454..637G},  the spin orientation of galaxies \citep[e.g.][]{2010MNRAS.405..274H} and their ellipticity \citep[e.g.][]{2015MNRAS.454.3341C}. 
On the other hand, other works suggest that the star formation rate is more dependent on the stellar mass and not on the environment \citep[e.g.][]{2012MNRAS.423.3679W}.  More recently, \citet{alp15} observed that  the large-scale environment of 
galaxies determines their stellar mass function,  but has otherwise a modest impact on other galactic properties. 
An important motivation to study the filamentary structure of the cosmic web as well as the simulated galaxies in filaments is to understand which internal galactic properties are connected to the large-scale properties of filaments, and which are independent.
We have studied in detail the properties of simulated dark-matter halos hosting galaxies in filaments, and compared them with those of real galaxies in the GAMA survey \citep[][]{driver09,driver11,liske15}.\\

Many high-resolution N-body cosmological simulations have been
used to explore the properties of the cosmic web \citep[e.g.][]{2005MNRAS.359..272C,2007MNRAS.381...41H,2012MNRAS.421L.137L,2014MNRAS.441.2923C}. Most of these works, however, 
focus on the properties of the dark component, described via N-body algorithms, representing
matter as a set of massive particles. In this work, we focus instead on the gas component,
described by our code of choice, ENZO \citep{enzo13} as an Eulerian fluid on a computational
mesh, which, for our purposes, has been kept at fixed resolution.
We have exploited the fluid and thermodynamical properties of the gas to identify, extract and analyse 
the filamentary structures from the results of state-of-the art cosmological magneto-hydrodynamical simulations \citep[][]{va14mhd}.
Our methodology, presented in all details in \citet{gh15} (hereafter Paper I),
relies on the usage of the VisIt data visualization and analysis software \citep{Childs11visit:an},
exploiting a combination of its {\it Isovolume} and {\it Connected Components} algorithms.
The adopted numerical and data processing methodologies are described in Sec. \ref{sec:nummethods}.

In Sec. \ref{sec:statistics} and \ref{sec:galaxy}, we present the results obtained from several simulations of the $\Lambda$CDM cosmological model
with density parameters $\Omega_0 = 1.0$, $\Omega_{\rm BM} = 0.0455$, $\Omega_{\rm DM} =
0.2265$ (BM and DM indicating the baryonic and the dark matter respectively), 
$\Omega_{\Lambda} = 0.728$, and a Hubble constant $H_0 = 70.2$km/sec/Mpc. In particular, Sec. \ref{sec:statistics}
focuses on the study of the time evolution of the geometric, thermal and magnetic properties of filaments up to $z=1$. 
Sec. \ref{sec:galaxy} instead analyses the properties of the galactic halos hosted in filaments, in comparison to the 
results of the GAMA survey. 

The results are discussed in Sec. \ref{sec:discussion}, where we comment on the most important 
observational implications of our modelling of galaxies in filaments.  
The major achievements are finally summarized in Sec. \ref{sec:conclusions}.

\section{Numerical Methods}
\label{sec:nummethods}

\subsection{Numerical Simulations}
\label{sec:enzo}

The {\enzo} code
is an Adaptive Mesh Refinement (AMR) code designed to solve a broad variety of astrophysical problems. 
ENZO solves the equations describing the dynamics of the two main matter component in the
universe, dark and baryonic matter, driven by the gravitational
field generated by the combined mass distributions.
The dynamics of the collisionless DM component is followed using a particle-mesh N-body method, while the  
BM component is represented as an ideal fluid, discretised on a mesh.
The gravitational potential of BM+DM is calculated solving the Poisson equation through a combined multigrid plus Fast Fourier Transform approach.
The magnetic field in the magneto-hydrodynamical (MHD) method uses the conservative Dedner 
formulation of MHD equations (Dedner et al. 2002) which uses hyperbolic divergence cleaning to preserve  $\nabla \cdot \vec{B}=0$. 
In this work we do not use the adaptive mesh refinement (AMR) capability of ENZO. Although AMR can provide high spatial resolution, 
applied to filaments it would lead to excessive computational overhead, in particular in terms of memory required to store the AMR related data structures, due
to the need of refining large fractions of the computational volume. On the other hand, the resolution achieved with the regular mesh adopted in our runs  (see below)
is suitable  to properly describe the behaviour of moderate over-density structures, like filaments.
Further numerical details and resolution tests of these simulations are given in \citet[][]{va14mhd} and \citet[][]{va15radio}.

The simulations have been performed on the Piz Daint HPC system operated by the Swiss National Supercomputing Centre (ETHZ-CSCS),
a Cray XC30 supercomputer accounting for more than 5000 computing nodes, each equipped with an 8-core
Intel SandyBridge CPU (Intel Xeon E5-2670) and an NVIDIA Tesla K20X GPU \citep[][]{2015JPhCS.640a2058G}.

The simulations presented here belong to the larger CHRONOS++ suite of MHD simulations
produced using Piz Daint, designed to study the generation of extragalactic magnetic fields
under a variety of scenarios for magnetogenesis.
The three runs adopt a fixed resolution mesh. Two runs, hereafter called 
``BOX200'' and ``BOX100'', have a box size of (200 Mpc)$^3$ and (100 Mpc)$^3$ (comoving) respectively,
and a mesh of $1200^3$ cells. The third one, called ``BOX50'' simulates a box 
of (50 Mpc)$^3$ (comoving) with a mesh of $2400^3$ cells. The number of DM particles
is equal to the number of cells. The first two runs provide a large statistical sample and probe cosmic volumes of the
order of what  radio telescopes (e.g. LOFAR, SKA) can target with deep pointings. 
 However, these runs have a quite limited spatial resolution (around 167 and 83 kpc comoving).
The $2400^3$ run allows us to study in detail the properties 
of  several single objects at high spatial resolution (around 21 kpc comoving). 
The comparison between the three runs allows estimating the effect of numerical resolution and box size on the results. 
In all runs we initialised the magnetic field at the (comoving) value of $B_0=10^{-10}\rm G$, which is within the
present upper limit from the analysis of the CMB \citep[][]{ade15}.

\begin{table}
\caption{List of the simulations run for this project. First column: box length the simulated volume;  second column: number of grid cells in the initial conditions;  third column: spatial resolution; fourth column: DM mass resolution; fifth column: initial redshift: sixth column: initial comoving magnetic field.}
\centering \tabcolsep 5pt
\begin{tabular}{c|c|c|c|c|c}
 $L_{\rm box}$ & $N_{\rm grid}$ & $\Delta x$ & $m_{\rm DM}$ & $z_{\rm initial}$ & $B_{\rm 0}$ \\\hline
 [Mpc/$h$]  &              & [kpc/$h$]  &  $[M_{\odot}/h]$ & & [G]\\  \hline
 50 & $2400^3$ & $20.8$ & $9.8 \cdot 10^{5} $ & $49$ & $10^{-10}$\\
 100 & $1200^3$ & $83.3$ & $6.2 \cdot 10^{7} $ & $33.5$  & $10^{-10}$\\
 200 & $1200^3$ & $166.6$ & $4.9 \cdot 10^{8} $ &  $30.7$ & $10^{-10}$\\
 \end{tabular}
\label{tab:sim}
\end{table}

We assumed a WMAP 7-year cosmology with
$\Omega_0 = 1.0$, $\Omega_{\rm BM} = 0.0455$, $\Omega_{\rm DM} =
0.2265$, $\Omega_{\Lambda} = 0.728$, Hubble constant $H_0 = 70.2$km/sec/Mpc, a normalisation for the primordial density power
spectrum $\sigma_{8} = 0.81$ and a spectral index of $n_s=0.961$ for the primordial spectrum of initial matter
fluctuations \citep[][]{2011ApJS..192...18K}. The detailed list of the simulations analysed in this project is given in Tab. \ref{tab:sim}.

\subsection{Filaments Identification and Selection}
\label{sec:visit}

In order to identify filaments in the simulations, we have implemented a procedure 
based on the VisIt data visualisation and analysis framework \citep{Childs11visit:an}. A detailed description 
of such procedure, with validation tests and performance analysis, can be found in Paper I. 
Here we summarize its main features. 

The mass density of the baryonic component is used in order to separate over- from under-dense regions.
The {\it Isovolume}-based approach, described in \citet{Meredith2004}, has been adopted in order to 
accurately trace the boundaries between the warm/hot phase which characterizes collapsed cosmological structures 
and the cold under-dense phase typical of voids. Isovolumes are then segmented into distinct objects
applying a {\it Connected Components} filter.
Identified overdense isovolumes undergo a further cleaning procedure in order to eliminate objects that 
either cannot be classified as filaments or that can be affected by large numerical errors, due, in particular 
to the limited spatial resolution. The main steps of such procedure are:

\begin{enumerate}

\item {\it Identification and removal of large clumps.}
Galaxy clusters are identified using the Isovolume algorithm out of the highest peaks of the
matter distribution. They are discarded by removing from the data all cells identified as 
part of the cluster.

\item {\it Filament identification.}
The Isovolume algorithm is used once more on the residual cells to identify the volumes
with:
\begin{equation}
\delta_{\rm BM} = {\varrho_{\rm BM}\over\varrho_0} \ge a_{\rm fil}, \label{eq:BM}
\end{equation}
where $\delta_{\rm BM}$ is the BM over-density, 
$\varrho_{\rm BM}$ is the cell's baryon mass density, $a_{\rm fil}$ is a proper threshold
and $\varrho_0$ is the critical density.
The Connected Components algorithm is then used to combine cells belonging to distinct (i.e. non
intersecting) filaments, assigning to each filament an Id (an integer number) and marking each cell
belonging to a filament with that Id.

\item {\it Shape selection.}
We use the following two-stage cleaning procedure to remove possible round-shaped, isolated
structures that could be still present in the identified objects. 
First, we retain only elongated objects, with:
\begin{equation}
{\rm MAX}(r_{xy}, r_{xz}, r_{yz}) > \alpha ,
\end{equation}
where $r_{ab}$ is the ratio between axes $a$ and $b$ of the bounding box and $\alpha$ was set to 2.
Second, among the remaining objects, we accept as filaments all those whose volume is smaller than a given 
fraction of the corresponding bounding box volume (see Paper I for details).

\item {\it Data export.}
The properties of the cells belonging to each filament,
such as the mass density, temperature, velocity, geometric and topological information,
are exported to output files.

\end{enumerate}

The resulting methodology depends on six parameters, the most important being the mass density threshold
$a_{\rm fil}$, whose value was determined in Paper I to lie  in the range $0.5 - 2.0$. The exact value can influence some
statistical properties of the identified objects (e.g. the maximum size of filaments extracted from a simulation).
Other properties (e.g. the average density within filaments) are unaffected. Three of the remaining parameters 
are related to the physical properties of galaxy clusters and to the spatial resolution of the simulations (so they can
be considered as fixed for each model). The last two parameters have been set after extensive testing as discussed in Paper I 
(Appendix A). 

The whole procedure is implemented within the VisIt framework, which provides all the required
numerical methods and supports big data processing through a combination of optimized algorithms, the
exploitation of high-performance computing architectures, in particular
through parallel processing, and the support of client-server capabilities.
The VisIt filament reconstruction pipeline run on the Piz Dora CRAY XC40 system at CSCS, 
equipped with 1256 compute nodes, each with two 12-core Intel Haswell CPUs (Intel Xeon E5-2690 v3), using 
192 cores (this minimum number being set by memory constraints). 

\section{Properties of Filaments}
\label{sec:statistics}

In this section we present the statistical properties of the filaments  extracted
from the three different simulations. Results are presented at three redshifts, z=0.0, z=0.5 and z=1.0.

\begin{figure}
  \includegraphics[width=0.47\textwidth]{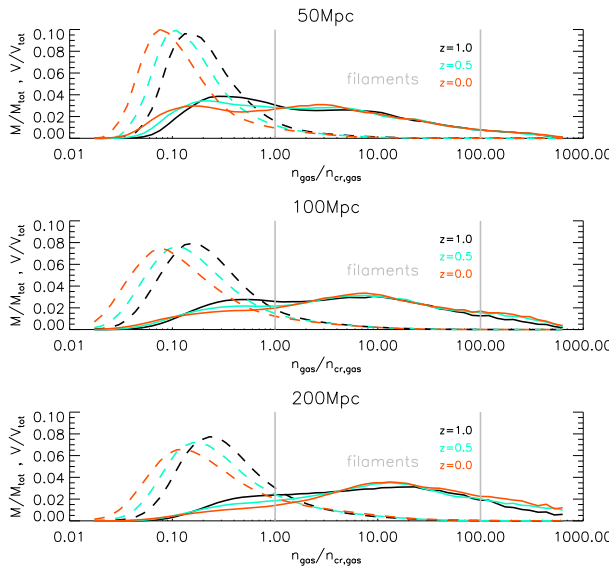}
  \caption{Mass (solid lines) and volume (dashed lines) fractions of the cells as a function of their BM over-density, for the three simulated volumes at three redshifts.}
  \label{fig:mvcell}
\end{figure}

\subsection{Global properties}
Before focusing on the properties of  filaments we analyse the mass and volume distribution of cosmic baryons as a function of the gas over-density, in three boxes at the different times, as shown in  Fig. \ref{fig:mvcell}.
While the volume fraction (dashed lines) is dominated by underdense regions, the mass distribution (solid) peaks within the typical
over-density range of filaments ($1 \leq \delta_{\rm BM} \leq 10^2$). With decreasing redshift more gas mass enters this over-density range, but the late growth
of massive galaxy clusters (more visible in the largest boxes) causes the distribution to move to higher over densities.

\begin{figure}
  \includegraphics[width=0.47\textwidth]{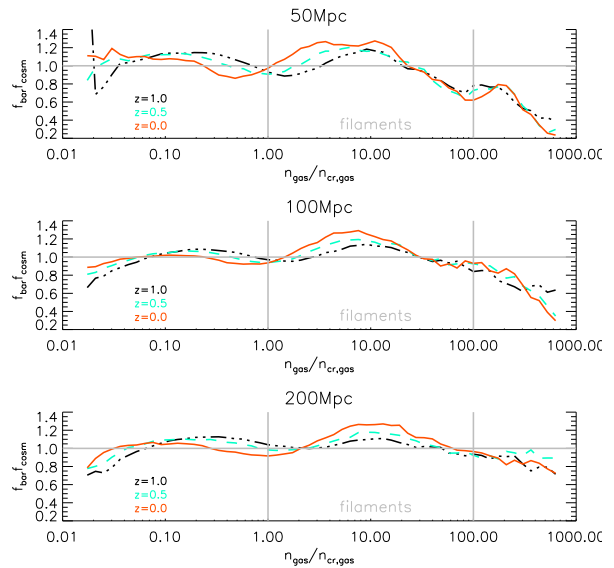}
  \caption{Baryon fraction of the cells as a function of their BM over-density, for the same dataset of Fig. \ref{fig:mvcell}.}
  \label{fig:bfraccell}
\end{figure}

In Fig. \ref{fig:bfraccell}, we show the baryon fraction for the same boxes and epochs, calculated as a ratio between the total BM mass and the total BM+DM mass within each over-density bin, normalised 
to the cosmic baryon fraction $f_{\rm cosmic} = 0.167$.
At the mean over-density of filaments, $\delta_{\rm BM} \sim 5-10$ (e.g. Paper I) the baryon fraction is larger than the cosmic one, consistent with the expectation that the WHIM dominates the
pool of ``missing baryons" \citep[][]{1999ApJ...514....1C,2001ApJ...552..473D,2015Natur.528..105E}. However, we shall see in the following that the baryon fraction enclosed within single filaments can be biased low by overdense substructures with baryon fraction $\leq f_{\rm cosmic}$. 
It is interesting to notice that the baryon fraction tends to be below the cosmic value even for overdensities around unity. 
This feature tends to drift to lower $\delta_{\rm BM}$ at lower redshifts, following the propagation of shock waves toward regions of decreasing
density.
Increasing the resolution, these features are enhanced and slightly move to higher $\delta_{\rm BM}$,
due to the impact of numerical diffusion, which keeps matter more diffuse around collapsing
objects.

\begin{figure*}
\centering
\begin{tabular}{|c|}
  \includegraphics[width=0.95\textwidth]{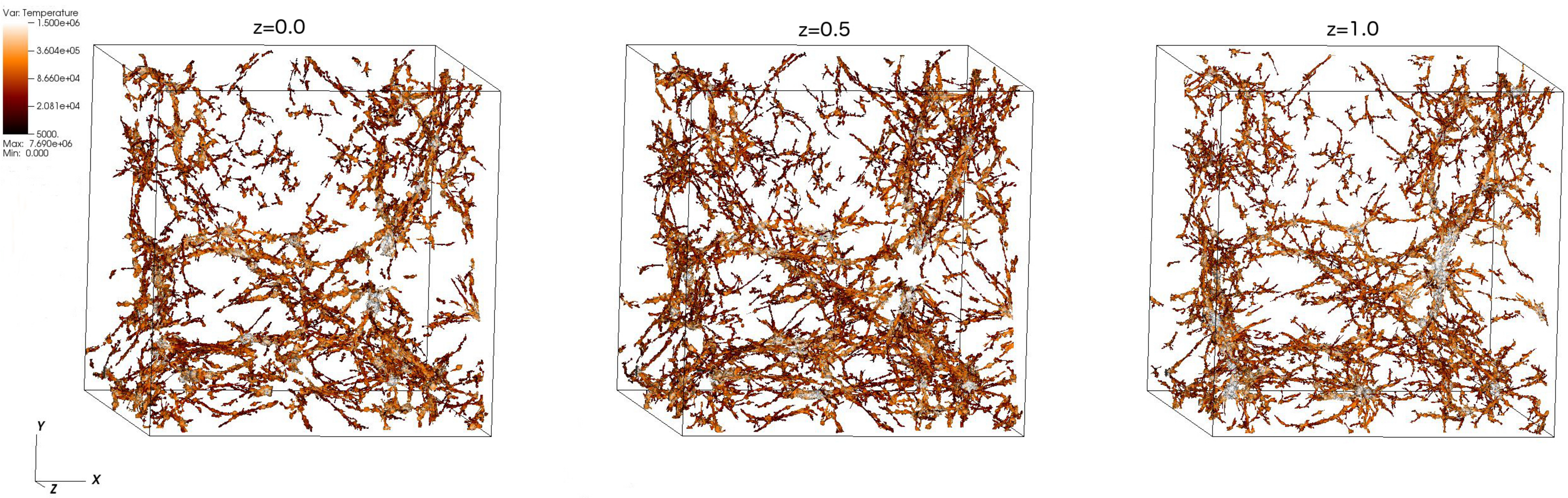} \\
  \includegraphics[width=0.95\textwidth]{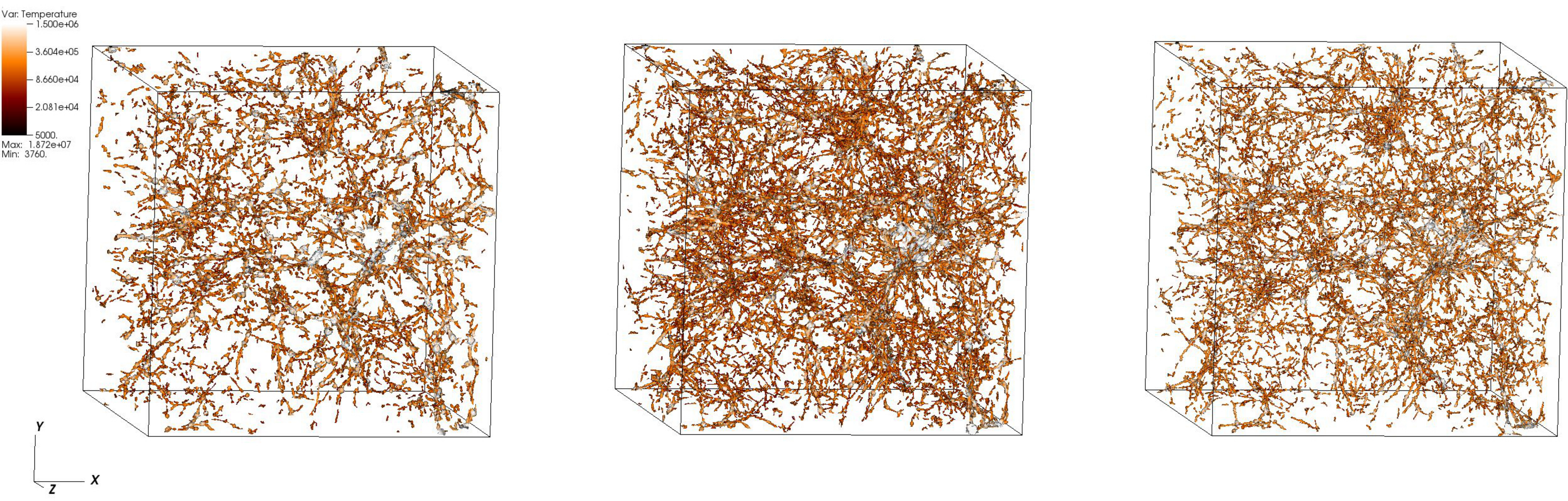} \\
  \includegraphics[width=0.95\textwidth]{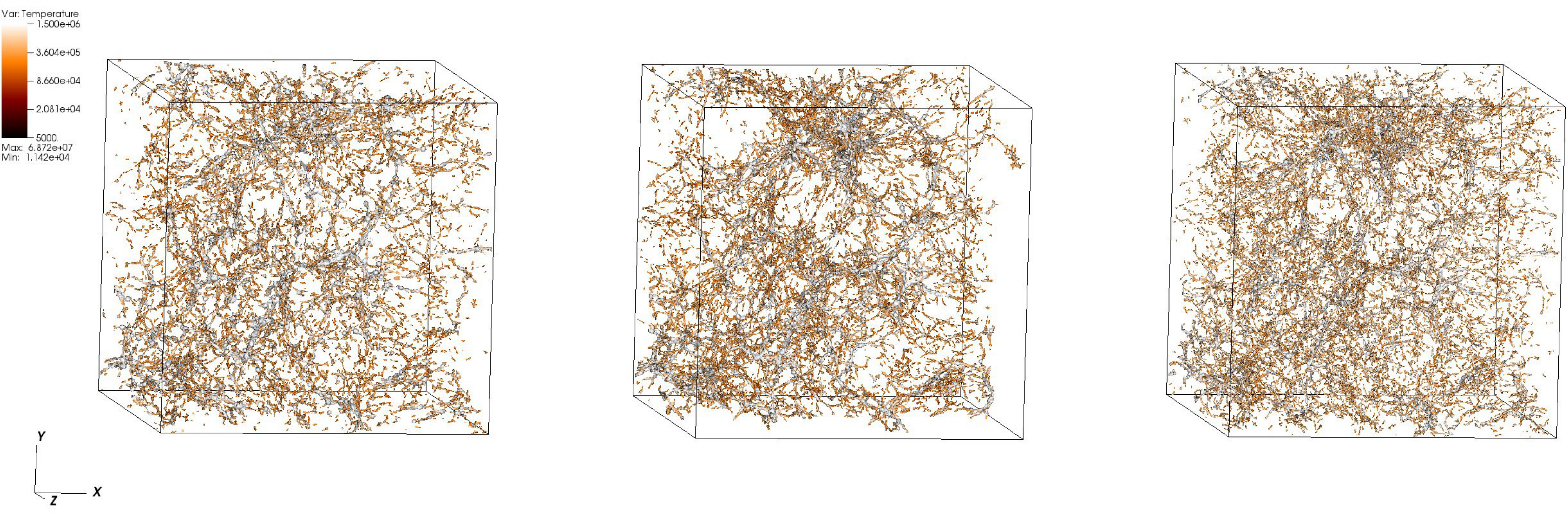} \\
\end{tabular}
\caption{Mass density isosurfaces for the three simulations at redshifts z=0.0 (left), z=0.5 (centre) and
z=1.0 (right). BOX50 is on the top, BOX100 in the middle, BOX200 is at the bottom. The entire simulation box is shown for all three runs.
Filaments are colored with their temperature. }
\label{fig:boxes}
\end{figure*}

Fig. \ref{fig:boxes} gives the visual impression of filaments, showing the isovolumes of over-density coloured according to the filament temperature. 
The filaments  encompass a broad range of length scales, from 
objects whose length is less than one Mpc,  
to structures whose linear size is comparable to that of the computational box (tens of Mpc).

Most filaments are already in place at $z=1.0$, and their evolution is not strong over the last $\sim 8 $ Gyr. 
However, the filamentary network gets slightly more volume filling from z=1.0 to z=0.5.  
At later times, filaments that were identified at earlier epochs can disappear due to being
absorbed into larger cluster structures progressively accreting matter from surrounding regions. 

\begin{figure*}
\centering
\begin{minipage}[c]{0.48\linewidth}
  \includegraphics[width=1.1\textwidth]{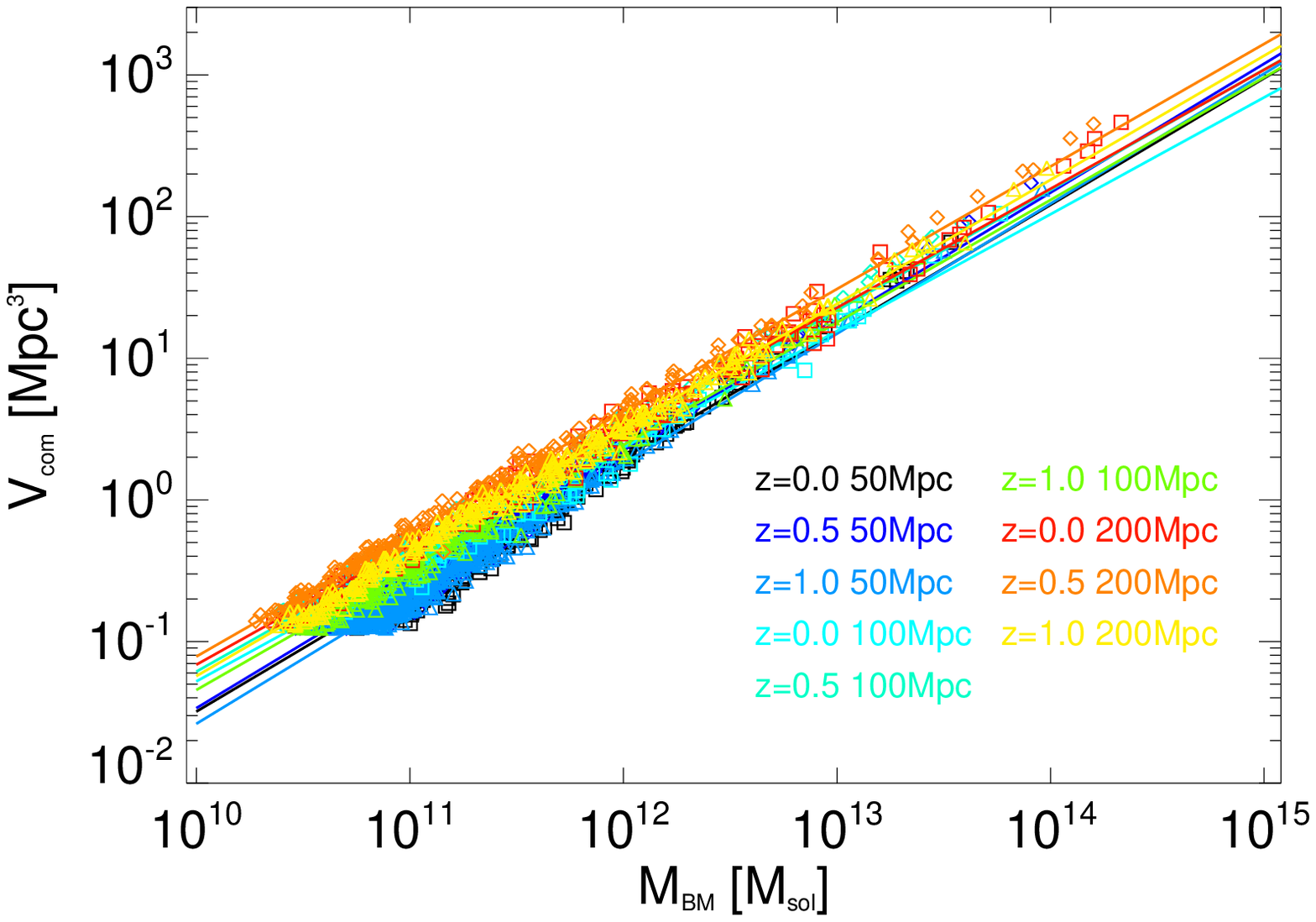}
\end{minipage}
\quad
\begin{minipage}[c]{0.48\linewidth}
  \includegraphics[width=1.1\textwidth]{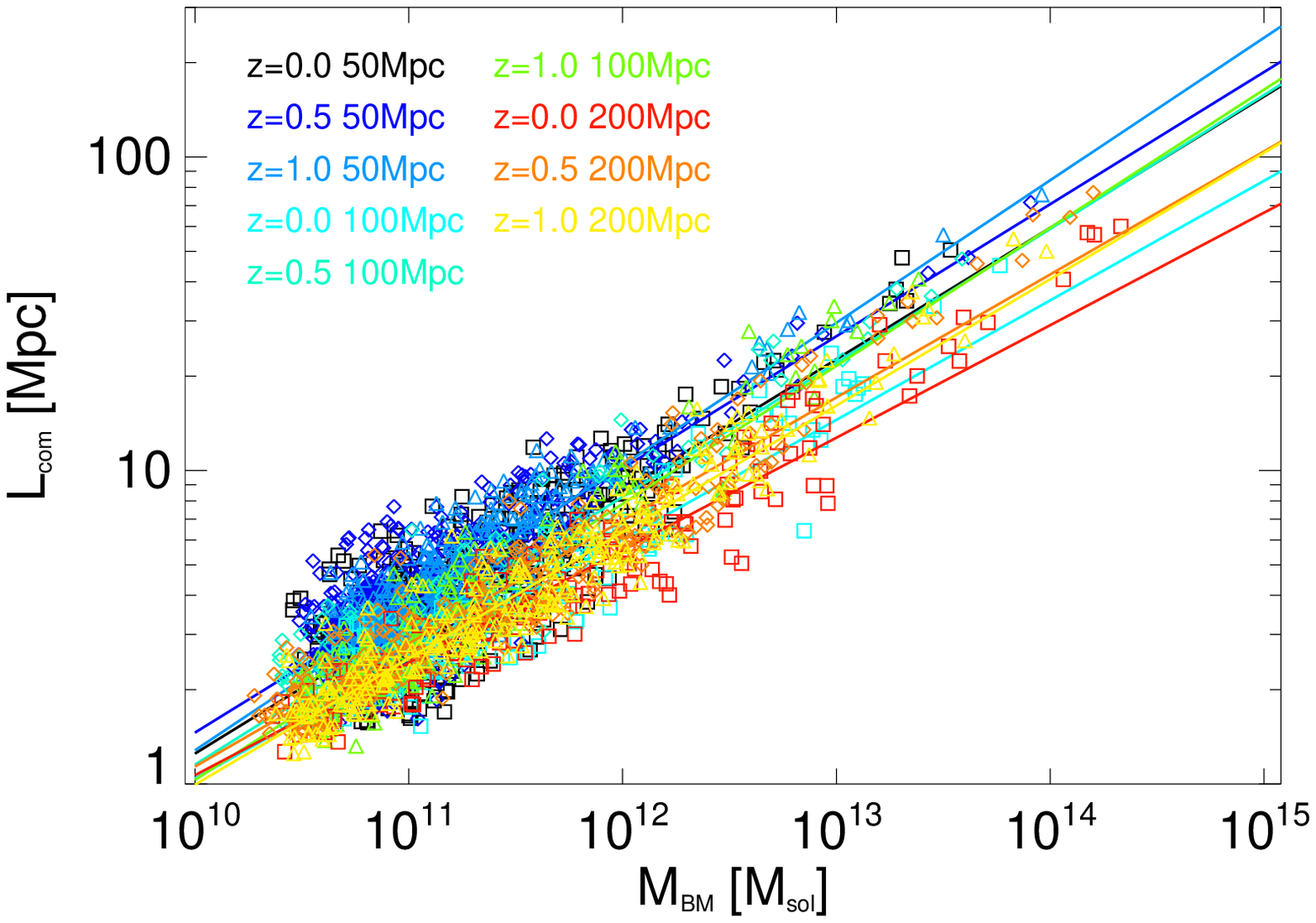}
\end{minipage}
\caption{Comoving volume (left panel) and length (right panel) of the filaments as a function of their BM mass. Squares represents 
data at z=0.0, diamonds at z=0.5 and triangles at z=1.0.}
\label{fig:vol}
\end{figure*}

\subsection{Geometric and Thermodynamical properties}
\label{subsec:thermo}
In Fig. \ref{fig:vol} we show the relation between the BM mass and the comoving volume/length of the extracted filaments, 
which are expected to be narrow power-law relations as found by \citet{2014MNRAS.441.2923C} and in Paper I. 
Tab. \ref{tab:fit} gives the values of the best-fit parameters for the log-log scaling relations, of the kind $Y=\beta M^{\alpha}$. 
Both, spatial resolution and redshift only slightly affect the slope. As expected the BOX200 simulation has the biggest filaments, as a result of the power on the largest scales, which is progressively reduced for BOX100 and BOX50.
At a given mass, filaments have smaller volumes but are longer in higher resolutions runs 
(a factor 2 to 5 when BOX50 and BOX200 are compared), due to the higher compression reached increasing the resolution.
On one hand, in fact, compression leads to higher mass densities, so largest masses in a given volume.  
On the other, objects with the same volume are thinner, so longer, at higher resolutions. 
(Fig. \ref{fig:mvcell}). Lengths at a given mass increase also at higher redshifts, due to the more diffuse distribution of matter. 
The slope and the normalisation of both relations show only a very mild evolution with redshift, showing the tendency that filaments with the same BM mass are slightly longer/larger at earlier times, consistent with the visual
impression of Fig. \ref{fig:vol}.

\begin{table*}
\caption{Best-fit parameters of the M-V, M-L and M-T relations. The $\alpha$ parameter is the exponent,
while $\beta$ is the normalization, i.e.  $Y=\beta M^{\alpha}$.}
\centering \tabcolsep 5pt
\begin{tabular}{|c|c|c|c|c|c|c|}
   \hline
   ID & \multicolumn{2}{c}{M-V} & 
   \multicolumn{2}{c}{M-L} & 
   \multicolumn{2}{c}{M-T} \\  \hline
   & $\alpha$ & $\beta$ & $\alpha$ & $\beta$ & $\alpha$ & $\beta$\\  \hline
  BOX50 z=0.0  & $0.894 \pm 0.009$ & $-10.437 \pm 0.103$ & $0.419 \pm 0.012$ & $-4.095 \pm 0.142$ &$0.447 \pm 0.016$ &$0.027 \pm 0.189 $\\
  BOX50 z=0.5  & $0.910 \pm 0.009$ & $-10.568 \pm 0.101$ & $0.421 \pm 0.013$ & $-4.050 \pm 0.144$ &$0.408 \pm 0.017$ &$0.456 \pm 0.189 $\\ 
  BOX50 z=1.0  & $0.918 \pm 0.007$ & $-10.760 \pm 0.088$ & $0.454 \pm 0.012$ & $-4.438 \pm 0.142$ &$0.384 \pm 0.017$ &$0.855 \pm 0.200 $\\
  BOX100 z=0.0 & $0.824 \pm 0.002$ & $-9.525 \pm 0.028$ &  $0.381 \pm 0.003$ & $-3.795 \pm 0.041$ &$0.431 \pm 0.005$ &$0.377 \pm 0.061 $\\
  BOX100 z=0.5 & $0.850 \pm 0.002$ & $-9.712 \pm 0.023$ &  $0.427 \pm 0.003$ & $-4.210 \pm 0.036$ &$0.375 \pm 0.004$ &$1.035 \pm 0.045 $\\
  BOX100 z=1.0 & $0.864 \pm 0.002$ & $-9.990 \pm 0.022$ &  $0.440 \pm 0.003$ & $-4.390 \pm 0.040$ &$0.375 \pm 0.004$ &$1.193 \pm 0.047 $\\
  BOX200 z=0.0 & $0.840 \pm 0.001$ & $-9.563 \pm 0.013$ &  $0.358 \pm 0.001$ & $-3.560 \pm 0.021$ &$0.370 \pm 0.003$ &$1.302 \pm 0.038 $\\
  BOX200 z=0.5 & $0.865 \pm 0.001$ & $-9.756 \pm 0.010$ &  $0.392 \pm 0.001$ & $-3.868 \pm 0.018$ &$0.344 \pm 0.002$ &$1.632 \pm 0.026 $\\
  BOX200 z=1.0 & $0.876 \pm 0.000$ & $-10.004 \pm 0.009$ & $0.403 \pm 0.001$ & $-4.032 \pm 0.018$ &$0.352 \pm 0.002$ &$1.680 \pm 0.024 $\\
\end{tabular}
\label{tab:fit}
\end{table*}

\begin{figure}
  \includegraphics[width=0.495\textwidth]{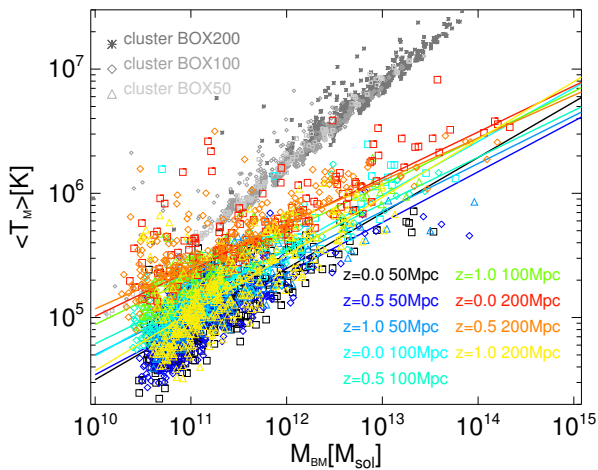}
  \caption{Average mass weighted filaments' temperature as a function of their BM mass. Squares represents 
data at z=0.0, diamonds at z=0.5 and triangles at z=1.0. The additional grey symbols show the distribution of galaxy clusters of the three boxes at $z=0$.}
  \label{fig:mt}
\end{figure}

The relation between the BM mass and the mass-weighted temperature is shown in Fig. \ref{fig:mt}. Resolution affects the scaling of the M-T relations
more than in the previous cases, filaments simulated at a lower resolution having a significantly
larger temperature in the same BM mass range. This follows from the tendency of accretion shocks to produce a larger thermalisation at coarser resolution, as known
from the study of cosmological shocks in the literature \citep[][]{va09shocks, va11comparison}. The scatter in this relation is also higher than in the previous cases, due to the
likely contamination from hot cluster outskirts in a few cases, as already discussed in Paper I (Sec 4.2). 
The slopes of the  $T \propto M_{\rm BM}^{\alpha}$  relation are in the range $0.35 \leq \alpha \leq 0.45$. This is significantly lower than the same relation for galaxy clusters, as is shown by the additional grey dots in Fig. \ref{fig:mt}. In the case of clusters, the slope is  $\alpha \sim 2/3$,  
as usually found in non-radiative simulations and is also expected from self-similar scaling relations \citep[e.g.][]{borgani08}. 
In clusters (dropping cosmological factors), the mass scales as $M \sim \varrho_c \Delta_c R_{\Delta}^3$, where $R_{\Delta} \sim M^{1/3}$, so that we get from $M kT \sim GM^2/R$  $T \propto M^2/3$ ($ \varrho_c$ being the critical density). For filaments this becomes: $M kT \sim GM^2/R_{\perp}$ and $M \sim \varrho_c R_{\perp}^2 \times L$, where $L$ is the length of the filament and $R_{\perp}$ the radius perpendicular to the axis. This then yields $T \propto M^1/2$ in case of cylindrical objects, where density only changes transverse to the cylinder axis.  

However, we expect  a significant departure from this simple geometrical argument because the
filaments have complex geometries. Furthermore, filaments with masses $M_{\rm BM}\sim10^{14} M_{\odot}$, are about 10 times cooler than galaxy clusters of the same mass, suggesting that
 the thermalisation of infall kinetic energy in filaments is about $\sim 10$ times less efficient than in clusters. This is consistent with the fact that filaments are not objects in virial equilibrium,  which sets them apart from galaxy clusters. 
 
\begin{table*}
\caption{Best-fit for the M-$f_{\rm BM}$ relation considering both all masses and only objects with
$M_{\rm BM} > 10^{13} M_{\odot}$. The $\alpha$ parameter is the slope and $\beta$ is the normalization. 
}
\centering \tabcolsep 5pt
\begin{tabular}{c|c|c|c|c|}
  ID & \multicolumn{2}{c}{All Masses} &
  \multicolumn{2}{c}{$M_{\rm BM} > 10^{13} M_{\odot}$} \\  \hline
  & $\alpha$ & $\beta$ & $\alpha$ & $\beta$ \\ \hline
  BOX50 z=0.0  &$0.026\pm 0.004$ &$0.672\pm 0.052 $ &$-0.001\pm 0.060 $ & $1.015 \pm 0.021 $\\
  BOX50 z=0.5  &$0.037\pm 0.004$ &$0.560\pm 0.054 $ &$-0.006\pm 0.002 $ & $1.023 \pm 0.001 $\\
  BOX50 z=1.0  &$0.027\pm 0.003$ &$0.663\pm 0.044 $ &$0.022 \pm 0.005 $ & $1.013 \pm 0.003 $\\
  BOX100 z=0.0 &$0.100\pm 0.004$ &$-0.295\pm 0.053$ &$0.020 \pm 0.013 $ & $0.976 \pm 0.005 $\\
  BOX100 z=0.5 &$0.109\pm 0.002$ &$-0.371\pm 0.031$ &$0.005 \pm 0.008 $ & $0.982 \pm 0.004 $\\
  BOX100 z=1.0 &$0.097\pm 0.002$ &$-0.234\pm 0.028$ &$0.018 \pm 0.009 $ & $0.973 \pm 0.004 $\\
  BOX200 z=0.0 &$0.107\pm 0.003$ &$-0.477\pm 0.046$ &$0.034 \pm 0.005 $ & $0.936 \pm 0.003 $\\
  BOX200 z=0.5 &$0.122\pm 0.002$ &$-0.640\pm 0.027$ &$0.026 \pm 0.004 $ & $0.940 \pm 0.002 $\\
  BOX200 z=1.0 &$0.114\pm 0.002$ &$-0.534\pm 0.023$ &$0.030 \pm 0.004 $ & $0.941 \pm 0.002 $\\
\end{tabular}
\label{tab:filfracfit}
\end{table*}

\begin{figure}
  \includegraphics[width=0.495\textwidth]{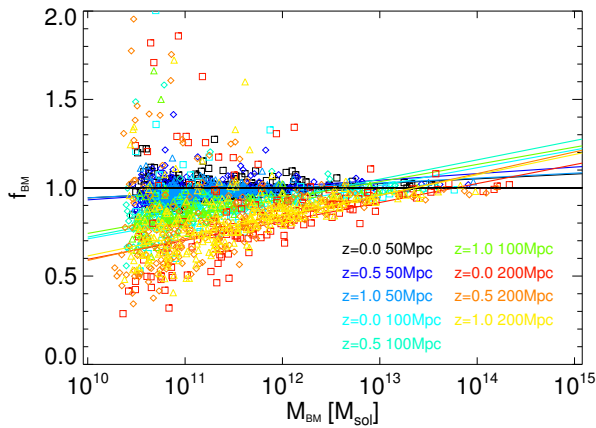}
  \caption{Baryon fraction in filaments as a function of their BM mass. Squares represents
data at z=0.0, diamonds at z=0.5 and triangles at z=1.0. The horizontal black line is the reference value $f_{\rm BM}=1$..}
  \label{fig:bfrac}
\end{figure}

The baryon fraction of each filament, $f_{\rm BM} = M_{\rm BM}/(M_{\rm TOT} \cdot f_{\rm cosmic})$ (where $M_{\rm BM}$ and $M_{\rm TOT}$ are the baryonic and total mass of the filament, respectively), is given in Fig. \ref{fig:bfrac}  in all boxes and for all epochs. The corresponding best-fit parameters are presented in Tab. \ref{tab:filfracfit}.
The baryon fraction in the largest filaments ($\geq 10^{13} M_{\odot}$) is very close to the cosmic value, while for lower masses 
it is from $10\%$ to $40\%$ lower than the cosmic value (the smallest values in BOX200). The highest baryon fraction at all redshifts is measured in the highest resolution run, which results in a very flat $f_{\rm BM} \propto M_{\rm BM}^{\alpha}$ relation, $\alpha \approx 0.05$.  At all resolutions there is no significant dependence on redshift. 
As already commented above (Fig. \ref{fig:bfraccell}), these results show that, although the WHIM gas in filaments has a baryon fraction above the cosmic mean, the total baryon faction locked within most identified filaments is biased low by the presence of high-density massive substructures with a lower baryon fraction. 

\subsection{Magnetic properties}
\label{subsec:magnetic}

\begin{figure}
  \includegraphics[width=0.49\textwidth]{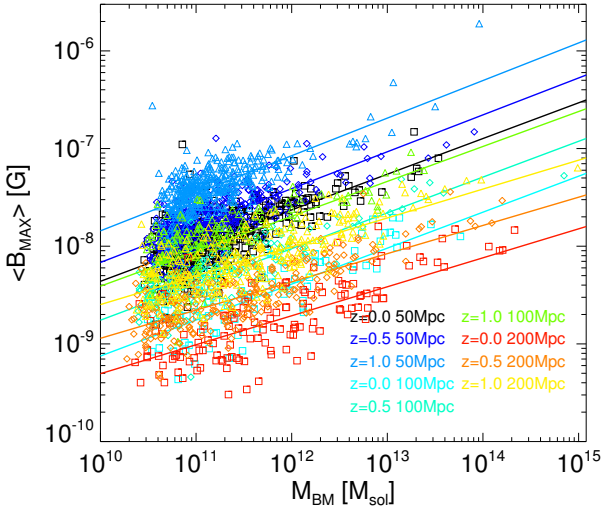}
  \caption{Distribution of the maximum absolute value of the magnetic field in filaments for the three simulated volumes at three epochs. }
  \label{fig:bfield_max}
\end{figure}

\begin{figure}
  \includegraphics[width=0.49\textwidth]{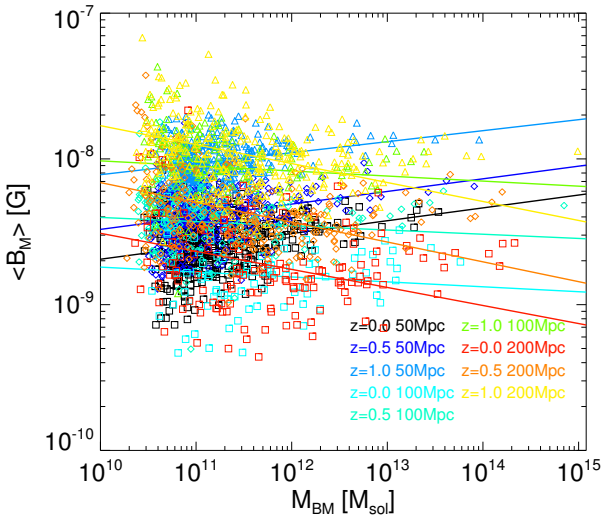}  
  \caption{Distribution of the mean volume-weighted magnetic field for each simulated filament for the three simulated volumes at three epochs. }
  \label{fig:bfield}
\end{figure}

The magnetisation level of cosmic gas outside galaxy clusters is largely unknown and the predictions from MHD simulations 
are still very approximate, due to
numerical limitation in resolution and physical uncertainties on the seeding processes \citep[e.g.][]{do08,donn09,va14mhd,2016MNRAS.456L..69M}. 
With our simulations we can study the average level of magnetisation in the cosmological seeding scenarios, and the dependence on the host filament mass. 
Fig. \ref{fig:bfield_max} shows the maximum physical magnetic field as a function of the filament gas mass, for the three volumes and epochs.
Compared to the thermodynamical quantities analysed above, the magnetic field in filaments shows the strongest evolution with resolution.
At each redshift, the maximum and the mean magnetic fields increase with resolution, as shown in Fig. \ref{fig:bfield}.
In the most resolved BOX50 run the largest filaments show a maximum of $\sim 0.1~ \rm \mu G$ at $z=0$, 
suggesting some modest level of dynamo amplification that can boost the magnetic field by $\sim 10^3$ times 
in a few patches within such objects. The maximum magnetic field slowly increases with mass ($B_{\rm MAX} \propto M_{\rm BM}^{\alpha}$, 
with $\alpha \sim 1/3$) and with resolution, suggesting that this process relates to dynamo amplification,
since more massive and more resolved structures are expected to host larger Reynolds number flows \citep[e.g.][]{va14mhd}. 
 However, even in the largest objects the average magnetic field (as shown in Fig. \ref{fig:bfield}) only show a growth of order $\sim 10$ compared to the comoving seed field, indicating that gas compression is on average the most significant source of magnetic field amplification in these objects. 
Overall, both the maxima and average physical magnetic fields are decreased by a factor $\sim 4$ from $z=0$ to $z=1$, consistent with the expected $B_{\rm phys} = B_{\rm com} (1+z)^2$ scaling (where $B_{\rm phys}$ denotes the physical fields strength and $B_{\rm com}$ the comoving field strength). This also indicates that the magnetisation of filaments has already reached its maximum at earlier epochs, and that no significant further amplification is found between $z=0$ and $z=1$, unlike galaxy clusters where the field is amplified over time due to the continuous stirring by turbulent motion (in agreement with \citet{va14mhd}). 
We conclude that unless other significant sources of magnetisation are present (e.g. magnetised winds from galaxies)
measuring the magnetic fields in filaments can constrain the amplitude 
of the initial primordial seed field. 

\subsection{Volumetric properties of filaments}
\label{subsec:profiles}

\begin{figure}
\centering
\begin{tabular}{c|c}
  \includegraphics[width=0.45\textwidth]{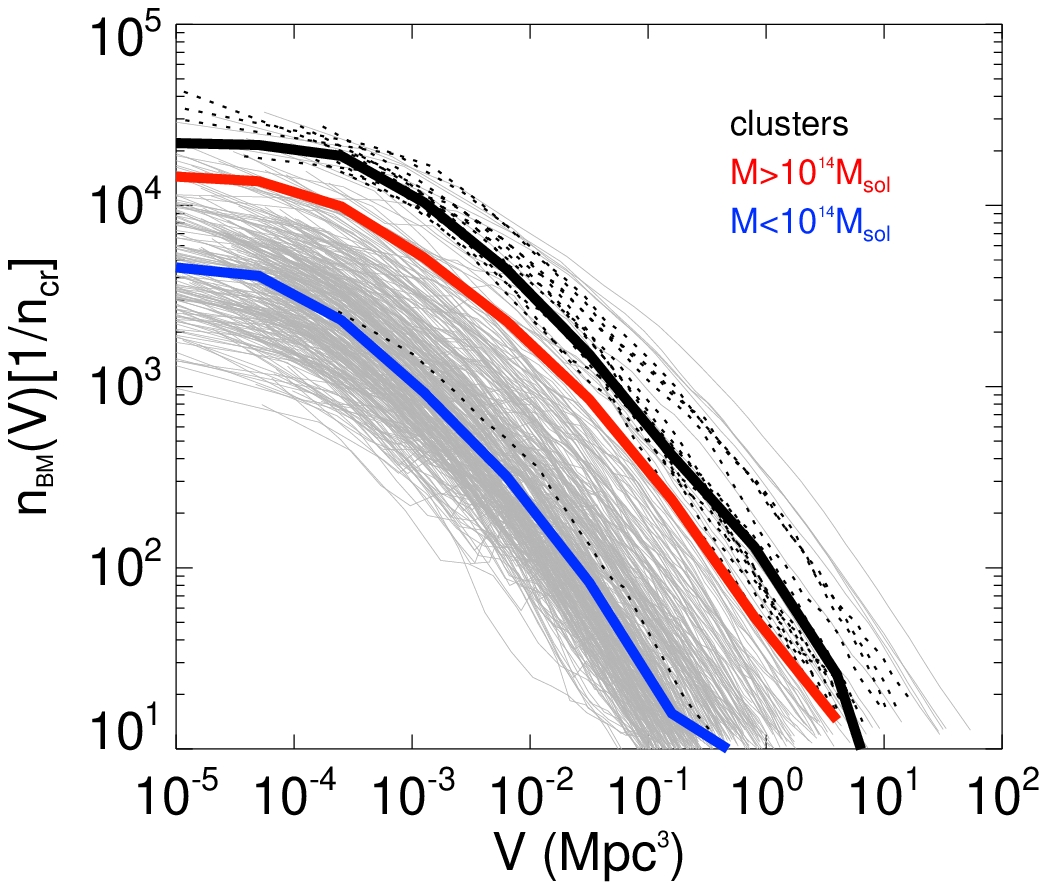}
  \end{tabular}
\caption{Gas over-density as function of volume for all filaments in the BOX50 run (grey lines) and for clusters 
(black dotted lines). The thick lines show the 
average distributions for the two mass selections of filaments, and for the clusters.}
\label{fig:prof_dens}
\end{figure}

The mean values of most quantities are shown here as a function of the gas over-density, motivated by the fact that
the average density in filaments decreases away from the filament axis. This was shown in Paper I, where we analysed a few filaments with regular shapes and computed radial profiles perpendicular to the filament axis. Here, we also show the volume distribution of the gas density and focus on the most resolved volume, BOX50. 
Fig. \ref{fig:prof_dens} shows the volume-density relation. 
The averaged profiles show a regular behaviour going from low-mass filaments ($M_{\rm BM} \leq 10^{14} M_{\odot}$, blue) 
to high-mass filaments ($M_{\rm BM} > 10^{14} M_{\odot}$, red) to clusters (black), showing that self-gravity causes a similar density stratification of 
gas in the volume of these structures.   In low-mass objects densities increasd slightly more gently towards smaller volumes, due to 
the combined effect of physical pressures and resolution effects. 

\begin{figure*}
\centering
\begin{tabular}{c|c}
  \includegraphics[width=0.45\textwidth]{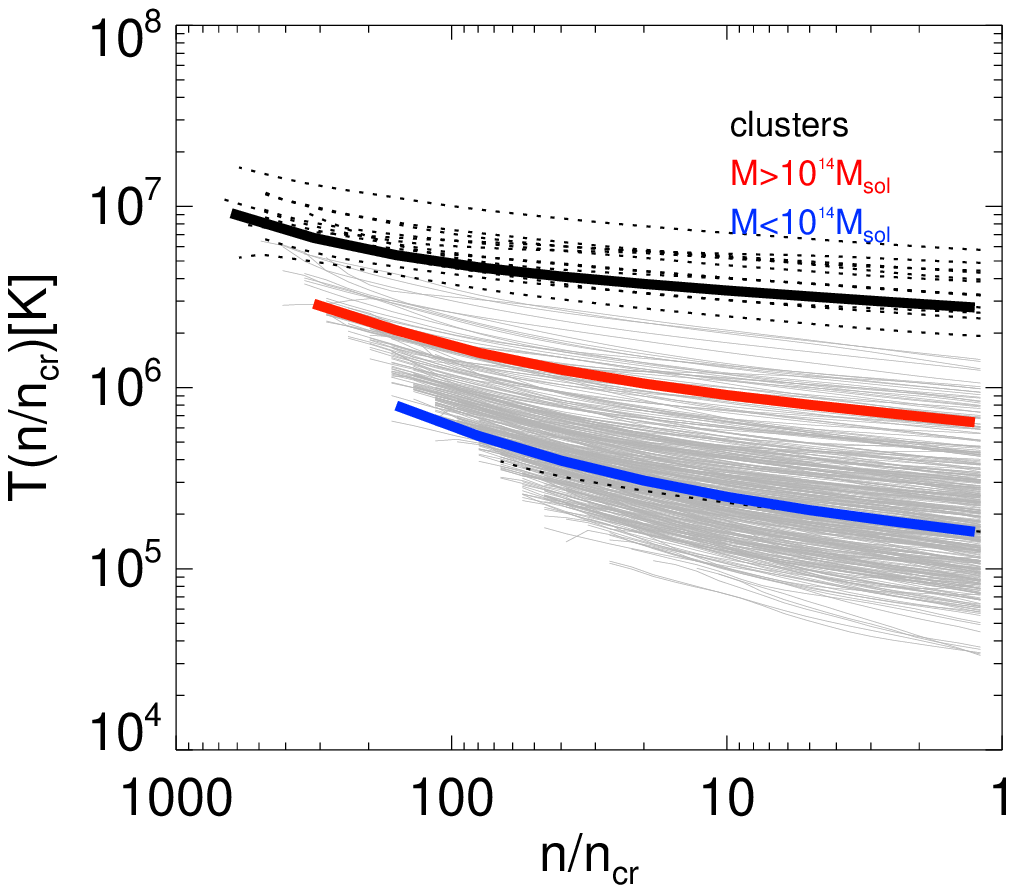} &
  \includegraphics[width=0.45\textwidth]{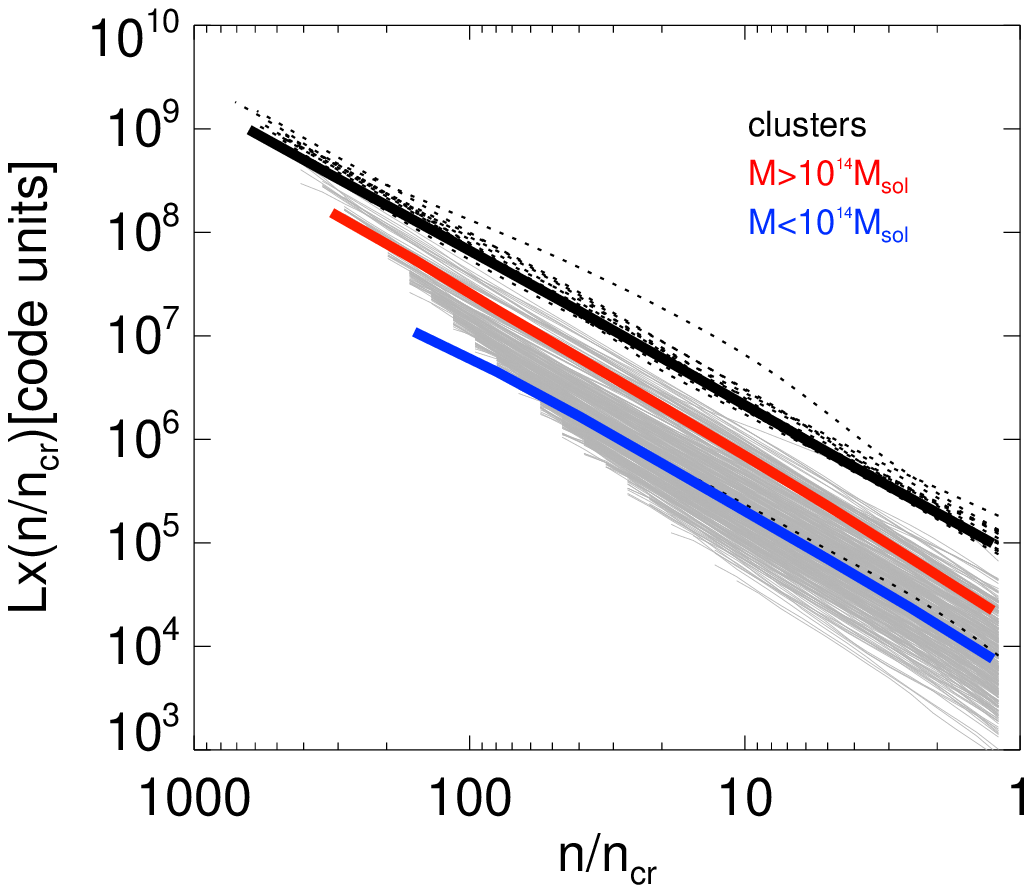} \\
    \includegraphics[width=0.45\textwidth]{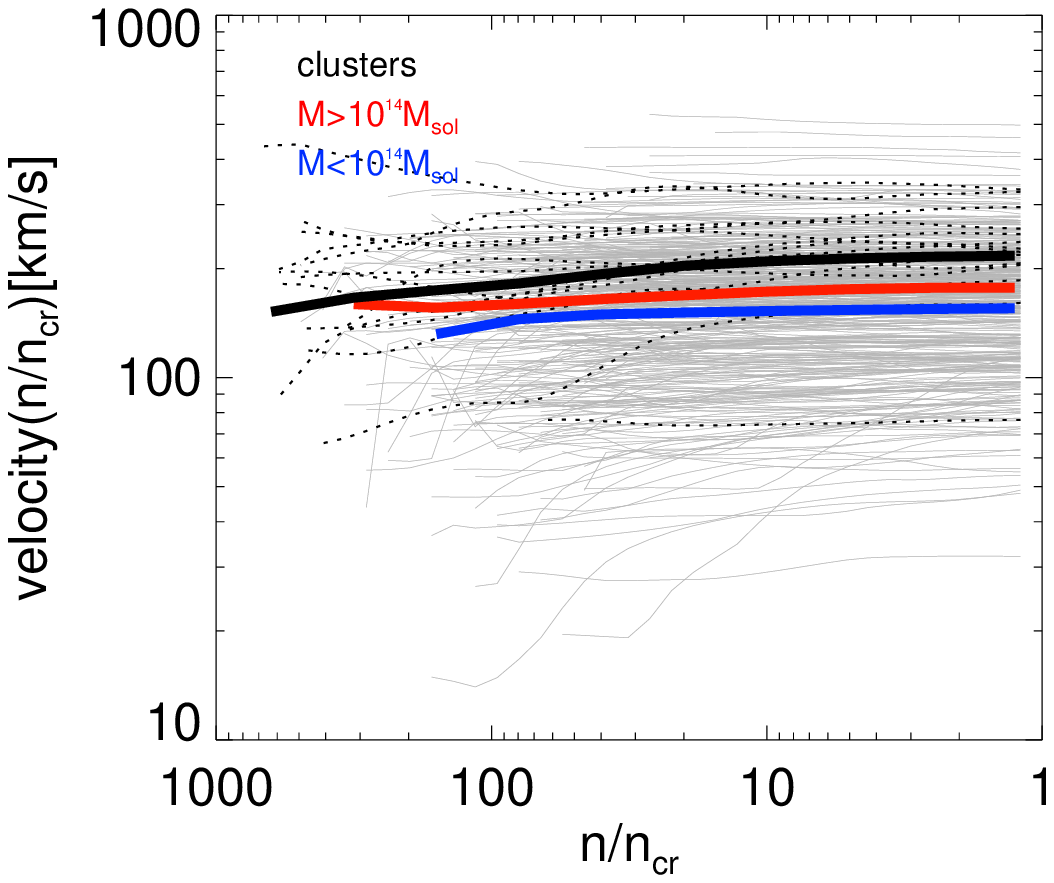} &
      \includegraphics[width=0.45\textwidth]{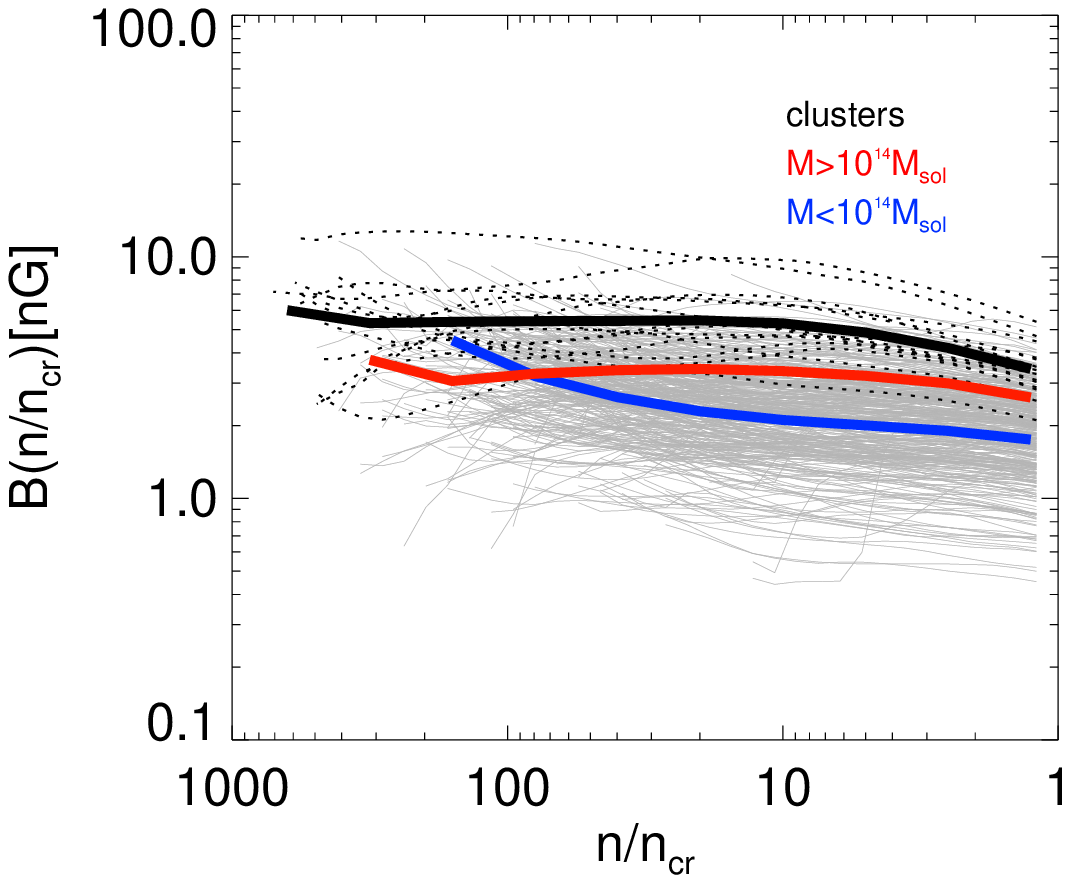}  
  \end{tabular}
\caption{Average temperature (top left), X-ray bolometric luminosity (top right), average velocity modulus (bottom left) and average magnetic field strength (bottom right) for all filaments in the BOX50 run. We show each filament individually with a grey line, and each cluster with a dotted black line. The coloured thick lines show the average relation for low mass (blue) and high mass (red) filaments, while the black thick line shows the average relation for clusters.}
\label{fig:prof_others}
\end{figure*}

In Fig. \ref{fig:prof_others} we show the average profile of temperature, bolometric X-ray emission ($L_{\rm X} \propto n^2 T^{1/2}$), average rms velocity and average magnetic field strength for the same objects, as a function of gas over-density. All the quantities show a clear trend with mass since smaller masses correspond to smaller volumes.  At the reference value of $n=50 n_{\rm cr}$ (roughly corresponding to the gas over-density at the virial radius of clusters), filaments are $\sim 10$ times cooler and less X-ray bright than clusters, while smaller filaments are up to $\sim 10^2$ times cooler and dimmer. 
This is expected because galaxy clusters produce the most energetic shocks, with temperatures higher than those of filaments 
at all density bins. The distribution displays remarkably similar shapes, stressing once more that self-gravity imposes a similar stratifications to these objects along the  direction in which density increases. 
On the other hand, smaller differences are seen in velocity and magnetic fields, and we find only  $\sim 2-3$ times lower typical values in filaments.  
The flat average velocity of $\sim 200 ~\rm km/s$ in the case of clusters corresponds to subsonic motions ($\mathcal{M} \sim 0.6 $ at the mean temperature of $\sim 5 \cdot 10^7$ K), while in the case of filaments the average velocity of $\sim 150 ~\rm km/s$ corresponds to transonic motions ($\mathcal{M} \sim 1 $) in large
mass filaments and to supersonic motions ($\mathcal{M} \sim 3$) in the smallest filaments. 
The predominance of supersonic compressive motions in filaments explains why no dynamo amplification is observed within them, consistent with our results in \citet{va14mhd}. At higher resolution in clusters, instead, the sub-sonic motions are expected to develop solenoidal turbulence and sustain a more significant dynamo amplification \citep[e.g.][]{ry08}, even if the clusters in BOX50 are not sufficiently resolved to develop a significant dynamo \citep[][]{va14mhd}. 

\section{Galaxies within filaments}
\label{sec:galaxy}

\begin{figure*}
\centering
\begin{tabular}{|c|c}
  \includegraphics[width=0.48\textwidth]{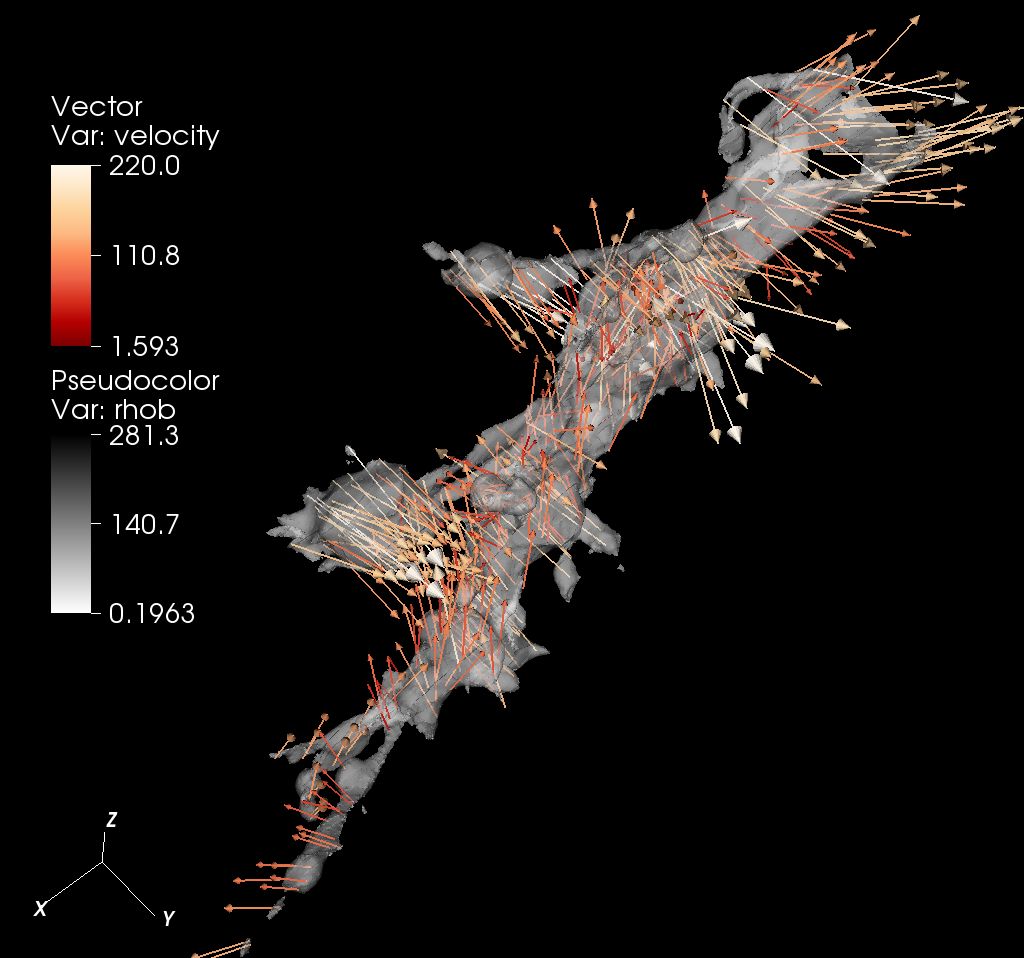} &
  \includegraphics[width=0.48\textwidth]{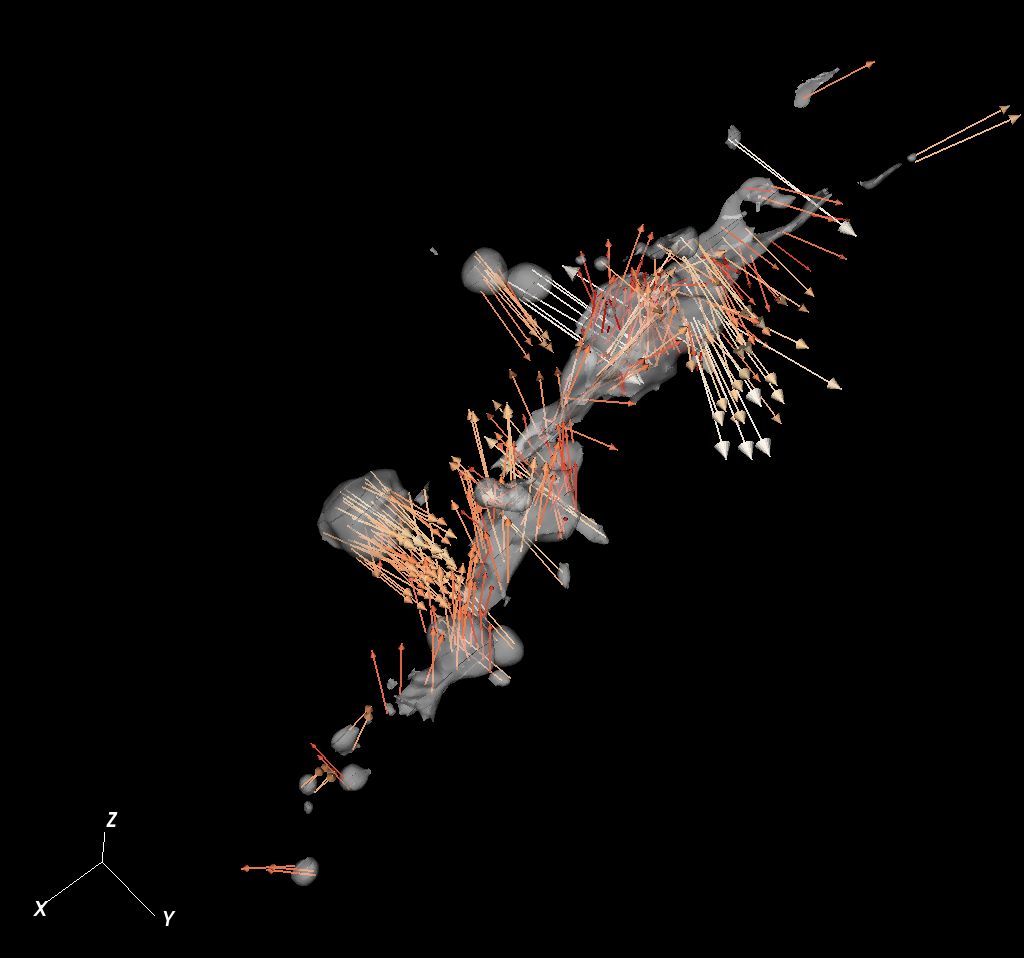} \\
  \includegraphics[width=0.48\textwidth]{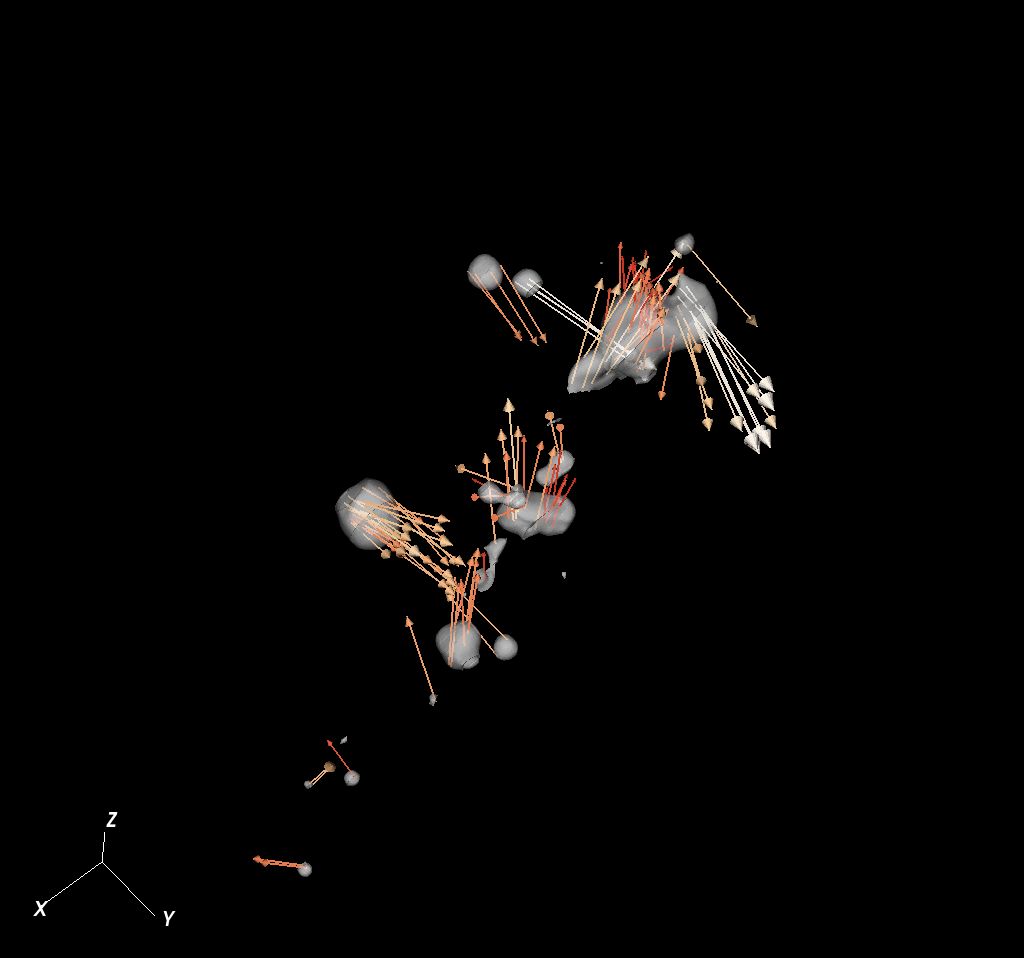} &
  \includegraphics[width=0.48\textwidth]{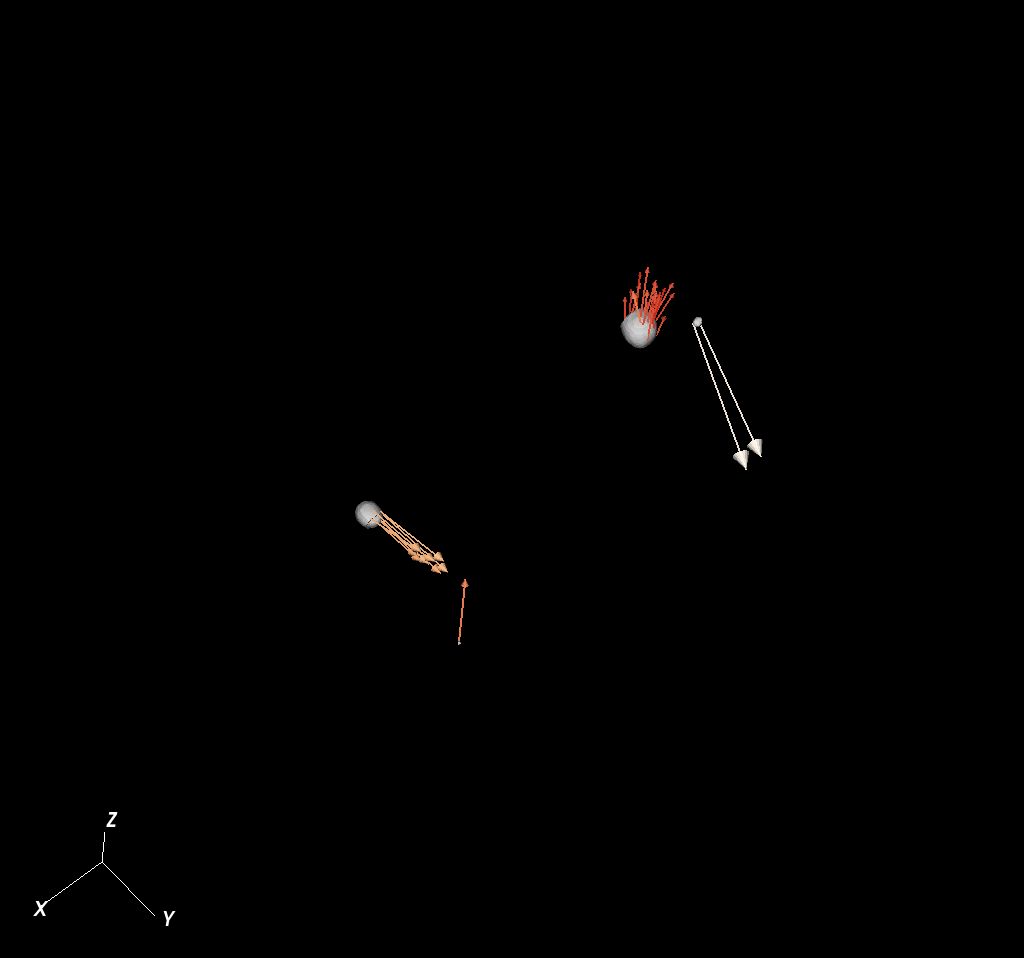} 
\end{tabular}
\caption{BM density isovolumes and velocity fields at different overdesities for filament 107,
extracted from BOX50 at z=0.0.
Isovolumes are calculated for over-density thresholds $a_{\rm fil}$ = 1.0 (top-left), 2.0 (top-right), 
5.0 (bottom-left) and 50.0 (bottom-right).}
\label{fig:rhoth}
\end{figure*}

\begin{figure*}
\centering
\begin{tabular}{|c|c}
  \includegraphics[width=0.48\textwidth]{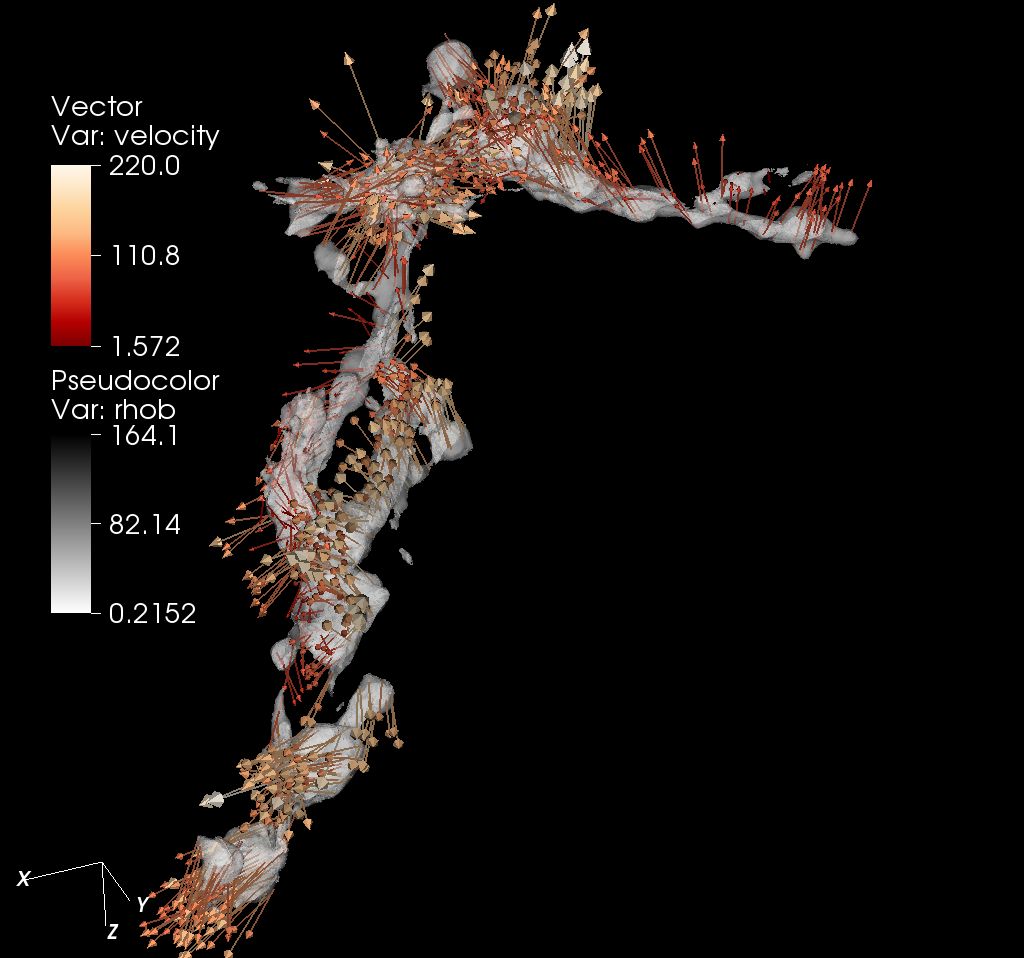} &
  \includegraphics[width=0.48\textwidth]{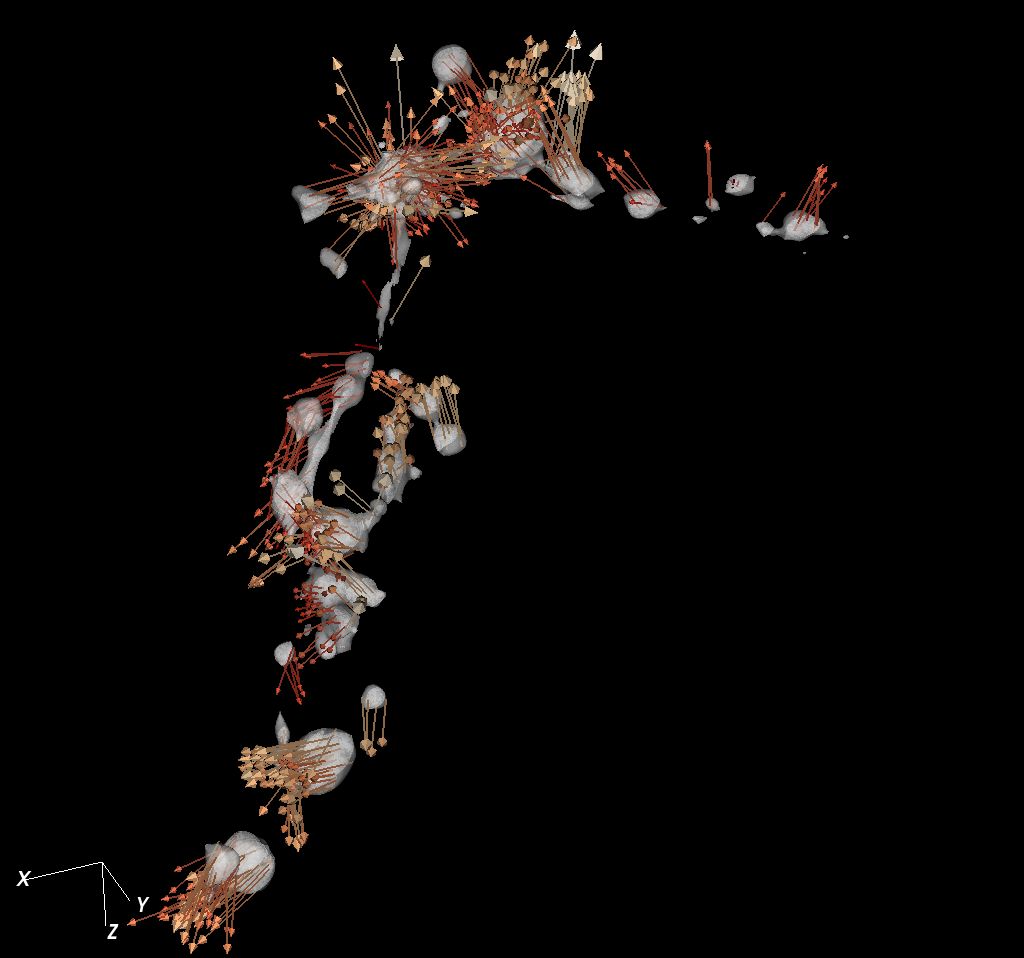} \\
  \includegraphics[width=0.48\textwidth]{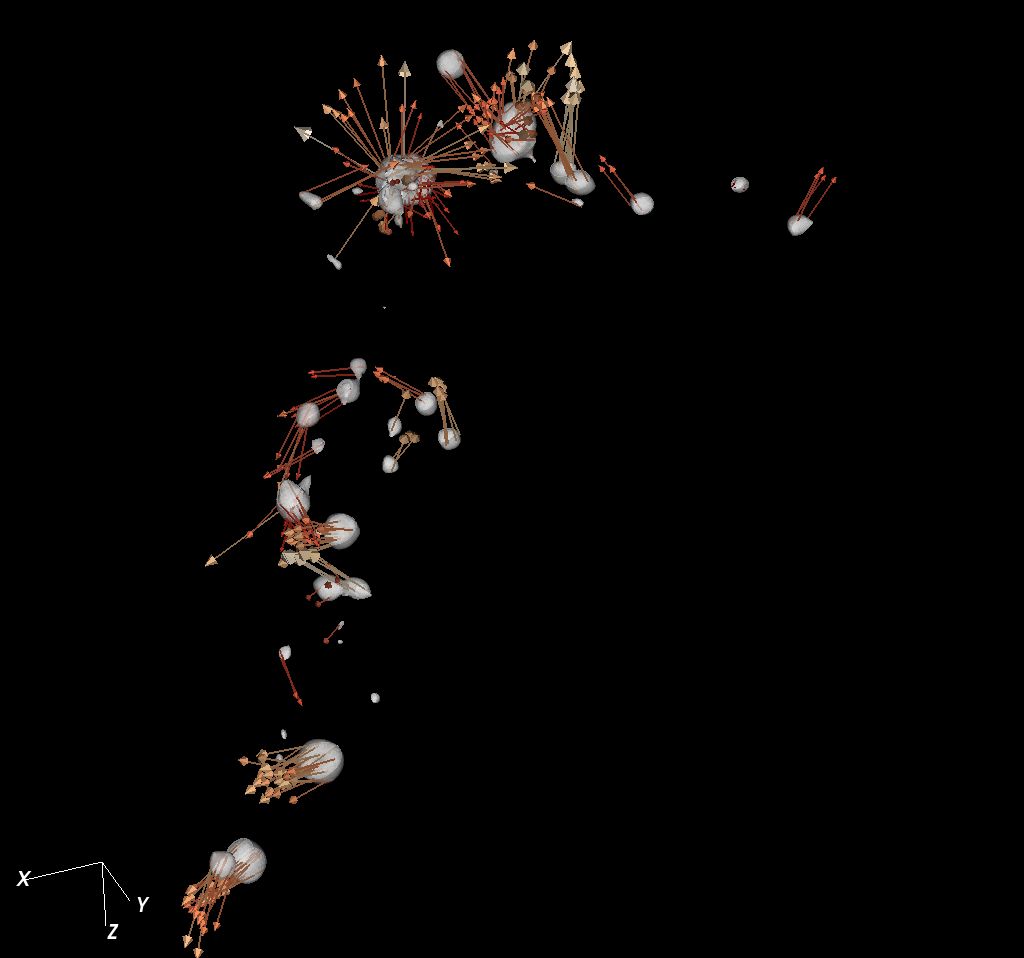} & 
  \includegraphics[width=0.48\textwidth]{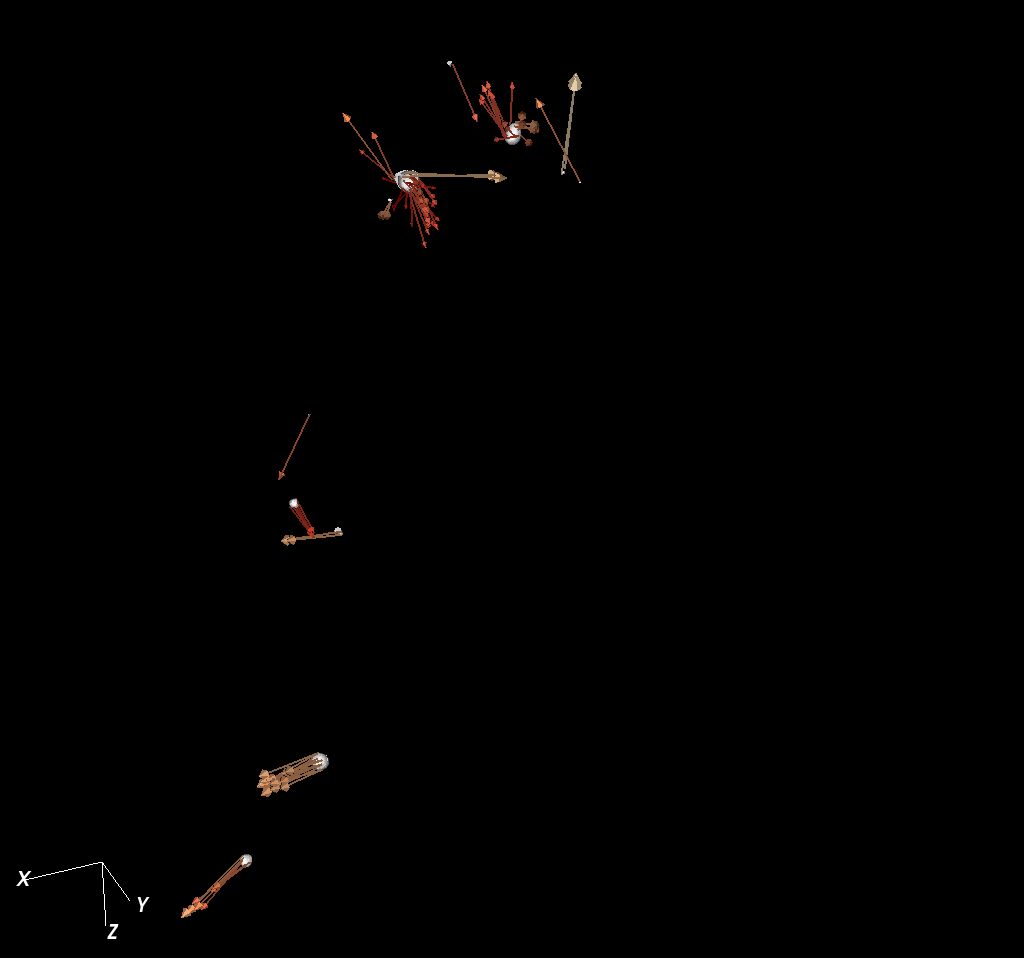}
\end{tabular}
\caption{BM density isovolumes and velocity fields at different overdesities for filament 10947, 
extracted from BOX50 at z=0.0.
Isovolumes are calculated for over-density thresholds $a_{\rm fil}$ = 1.0 (top-left), 2.0 (top-right), 
5.0 (bottom-left) and 50.0 (bottom-right).}
\label{fig:rhoth2}
\end{figure*}

\begin{table*}
\caption{Main physical properties of two sample filaments (ID 107 and 10947) extracted from BOX50 at z=0.0.
Columns contain (from left to right) the filament's ID, its length and volume, the BM and the total mass, the mass averaged temperature
and the average, minimum and maximumn magnetic field intensity.}
\centering \tabcolsep 5pt
\begin{tabular}{c|c|c|c|c|c|c|c|c|}
  ID    & L (Mpc) & V (Mpc)$^3$ & BM M (M$_\odot$) & Tot. M (M$_\odot$) & $<$T$>_M$(K) & $<$B$>_M$(G) & B$_{\rm min}$(G) & B$_{\rm max}$(G)\\  \hline
  107   & 10.36 & 2.71 & $1.58\times10^{12}$ & $9.16\times10^{12}$ & 586070.35 & $2.59\times10^{-9}$ & $1.45\times10^{-11}$ & $2.12\times10^{-8}$\\
  10947 & 13.25 & 3.16 & $1.91\times10^{12}$ & $1.14\times10^{13}$ & 342997.88 & $1.49\times10^{-9}$ & $0.41\times10^{-11}$ & $1.37\times10^{-8}$\\
\end{tabular}
\label{tab:filprop}
\end{table*}

Our selection procedure for filaments allows us also to analyse the over-dense objects they contain, which are tagged based on a density threshold, as explained in Sec. \ref{sec:visit}.
In Fig. \ref{fig:rhoth} and \ref{fig:rhoth2} we give 
two representative examples of how the appearance and structure of a filament change with the density threshold, $a_{\rm fil}$.
The reference value $a_{\rm fil}=1$ used to extract the filament properties of the previous sections is in the top-left panel.
The threshold increases from left to right and from top to bottom. Velocity vectors are superimposed. 
The main characteristics of the two objects, identified in our catalogue as filament 107 and 10947, can be found in Tab. \ref{tab:filprop}.
Increasing the value of the threshold, the extended filamentary structure changes progressively
into increasingly smaller and higher density spherical blobs, distributed along the spine of the filament.
The velocity fields show a rather complex distributions at all scales, with rather random velocity distribution in the proximity
of high density peaks and more coherent motions towards the extremes of filaments, connecting to surrounding clusters.  The highest density
blobs close to the central axis of filaments display large velocity fields and are not at rest with respect to the surrounding filament volume. 

The distribution of the blobs 
mostly marks the location of self-gravitating halos identifying the location of forming galaxies. 
Although the missing baryon physics and low resolution do not allow us to simulate
galaxy formation, their statistical and overall physical properties are expected to be captured by our runs. In particular, the total mass is
dominated by the DM component, which is not affected by the physics of the gas.
This will be verified in this section by comparing the properties of over dense halos extracted within our filament those of real galaxies.\\

For each filament we have extracted the volumes with total (BM+DM) over-density above a given threshold.
Only BOX50 has been used, the other two boxes having a spatial resolution too low to cope with 
objects at the galactic size scale. 
In order to increase the statistics of objects here we included also a second simulated volume
of $(50 ~\rm Mpc)^3$ with $2400^3$ cells/particles (BOX50B), in all identical to BOX50 except for the different random initial conditions. 
One of the best observational datasets that can be used to compare with the statistics of galaxies in the cosmic web is
the Galaxy And Mass Assembly (GAMA) survey, which is a spectroscopic survey of $\sim 300,000$ galaxies to $r=19.8$ mag, across five independent fields for total area of $\sim 286 ~\rm deg^2$ \citep[][]{driver09,driver11,liske15}.
\citet{alp14a} and \citet{alp14b} presented a classification of large-scale structures in 3 GAMA fields, including 
the detection of filaments of groups and galaxies, as well as of smaller structures formed by individual galaxies on the peripheries
of filaments (``tendrils") and galaxies in voids. 
The method used to generate the GAMA Large Scale Structure Catalogue (GLSSC) is based on an adapted minimum spanning tree algorithm that is first run on galaxy groups to identify filaments and the galaxies that inhabit them. This population of galaxies is then removed from the data, and a second minimum spanning tree is used to identify tendril and void galaxies. The groups used to identify filaments are taken from the GAMA Group Catalogue \citep[][]{2011MNRAS.416.2640R}, which are themselves identified using a friends-of-friends algorithm that is run in both projected and comoving space, and calibrated on a set of mock galaxy catalogue that match GAMA's volume and luminosity function \citep[][]{2013MNRAS.429..556M}. 

\begin{figure}
\centering
\includegraphics[width=0.48\textwidth]{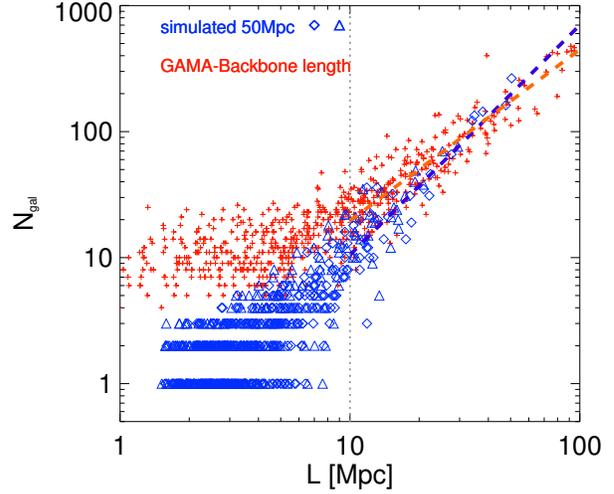}
\caption{Comparison of the number of galaxies detected in filaments of a given backbone length in the GAMA survey with the corresponding
number extracted from BOX50 and BOX50B runs. The vertical line marks the critical length outside which the two dataset show good agreement, while the coloured blue and red lines show the best fit for the log-log regression of both datasets.}
\label{fig:gama}
\end{figure}

The GLSSC catalogue consists of 
group centres, different size estimates,  luminosity and magnitude measurements etc. \\
The data of GLSSC are compared with those of the halos extracted within our simulated filaments by our procedure. 
In Fig. \ref{fig:gama} we present the number of galaxy objects as a function of the length of the host filaments. We notice that the lengths are here defined in slightly different ways due to the intrinsic differences in the datasets:  in our simulations it is the diagonal of the bounding box of each object, while in GAMA it is defined as the ``backbone length", defined as the total distance of the minimum spanning tree links from one end of the filament to the other \citep[][]{alp14a}. 
The blue diamonds represent the
objects extracted from the simulation (for an over-density threshold $a_{\rm gal} = 1000$), while the red crosses show the galaxies coming from the GAMA catalogue.  The number of observed and simulated galaxies is  consistent in filaments longer than $\sim 10$ Mpc
(corresponding to gas masses $\geq 10^{12}$M$_\odot$), and so is the log-log best-fit relations computed in these range of
length (see straight lines in Fig. \ref{fig:gama}). Based on these relations, a $\sim 10$ Mpc long filament hosts $\sim 10-30$ galaxies, while 
a $\sim 100$ Mpc filament hosts $\sim 400-600$ galaxies. 
In the case of filaments smaller than $\sim 10$ Mpc, both simulated and observed statistics are affected by systematic uncertainties
due to the slightly bigger impact of numerical effects (resolution, numerical pressure) on smaller objects on one hand,  
and to the adoption of the finite edge length which must be assumed in building the filament catalog \citep[][]{alp14a}, for GAMA data.

\begin{figure*}
\centering
  \includegraphics[width=0.945\textwidth]{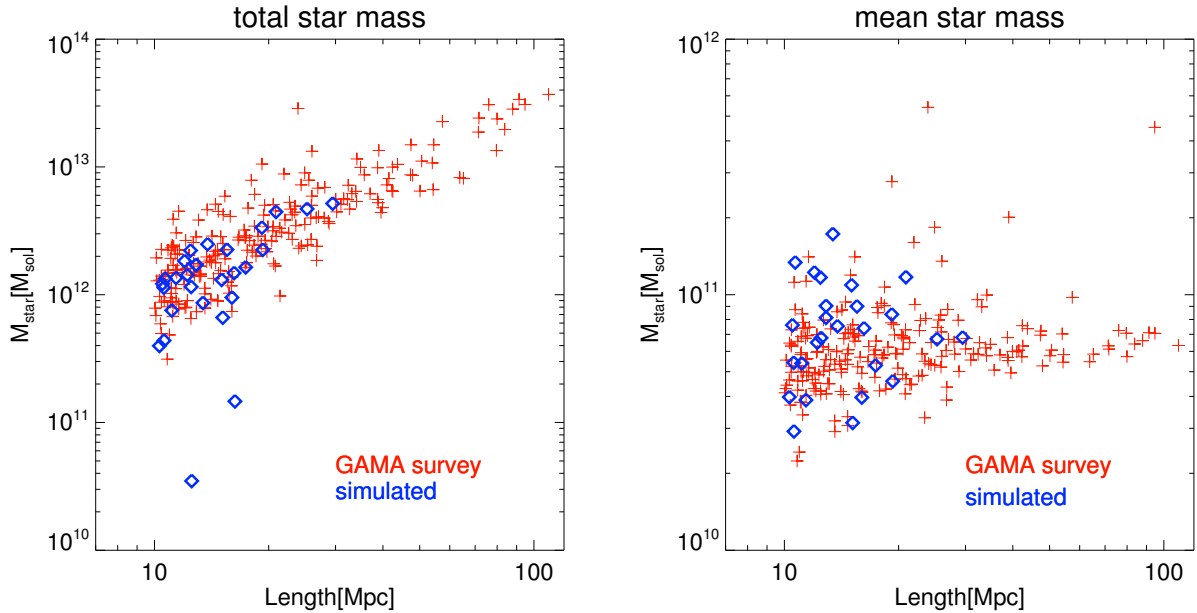}
\caption{Total stellar mass (left) and average stellar mass (right) of galaxies as a function of the length of the host filaments. The datapoints from 
GAMA \citet{alp14a} have been corrected for the finite aperture, while the stellar mass of simulated halos has been extrapolated using the formalism of \citet{gw11} (see text for details). In both cases the stellar masses are estimated within $10 R_e \approx 4 r_s$ (where $R_e$ is the half-light radius of the Sersic profile of each object and $r_s$ is the scale radius of the NFW profile).} 
\label{fig:stars}
\end{figure*}

Limited to $\geq 10$ Mpc filaments, we now show in Fig.~\ref{fig:stars} the total (left) and average (right) stellar mass in galaxies in the GLSSC dataset and 
in the extrapolation of our simulated galaxies. These runs lack the necessary physics to properly simulated galaxy formation (i.e. radiative gas cooling, 
star formation and star feedback) and therefore we can only extrapolate the stellar mass of our simulated halos based on other works. 
Similar to \citet{han15}, we assumed a scaling between the galaxy stellar mass and total mass within $R_{\rm 200}$ of the form:

\begin{equation}
M_* = AM_h [(M_h/M_0)^\alpha + (M_h/M_0)^\beta]^{- \gamma}
\label{eq:m*}
\end{equation}

with parameters $A=0.069$, $\log(M_0/h^{-1} M_{\odot})=11.40$, 
$\alpha=-0.926$, $\beta=0.261$ and $2.44$ as derived by \citet{gw11}.  

The stellar masses from galaxies in the GLSSC catalogue have been adjusted with a linear correction term as in \citet{taylor11}, which 
allows correction for the missing flux beyond the finite aperture used to model the observed spectral energy distribution.
In this way the stellar masses of all objects are referred to the extrapolated  total stellar mass in a Sersic profile truncated at $10 R_e$ ($R_e$ being the half-light radius). The predicted stellar masses from the simulated galaxies are accordingly computed up to $10 R_e \approx 4 r_s$ (where $r_s$ is the scale radius of the NFW profile of each halo). 
The comparisons in Fig. \ref{fig:stars} show that both datasets are in good agreement within the scatter, across one decade in the length of the host filament. Within the assumed start-to-mass scaling \citep[][]{han15}, these results suggest that both the average number of galaxies per filaments and the 
typical mass distribution of DM halos in simulations and observations are consistent.

\bigskip

\begin{figure*}
\centering
\begin{tabular}{c|c}
  \includegraphics[width=0.45\textwidth]{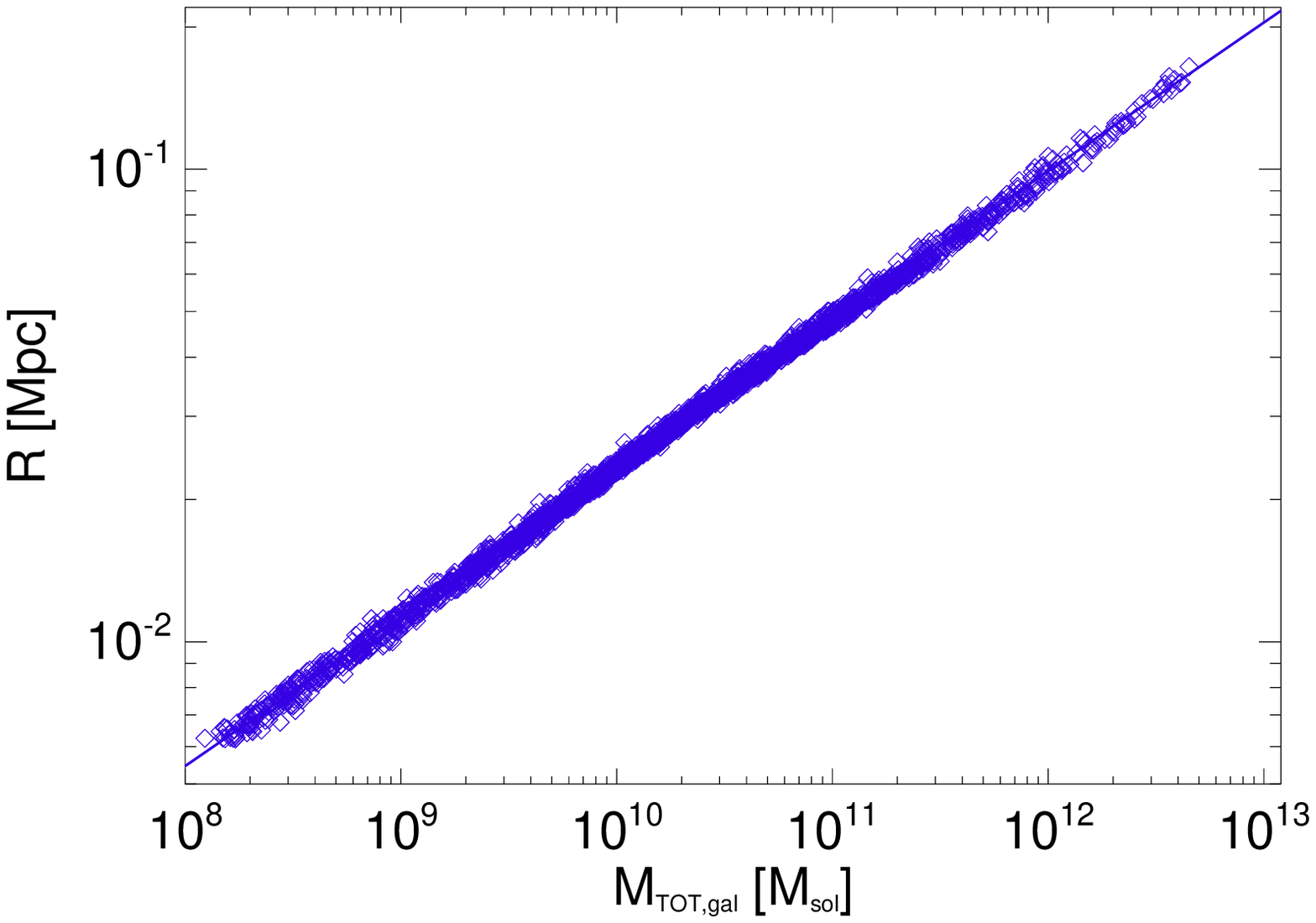} &
  \includegraphics[width=0.45\textwidth]{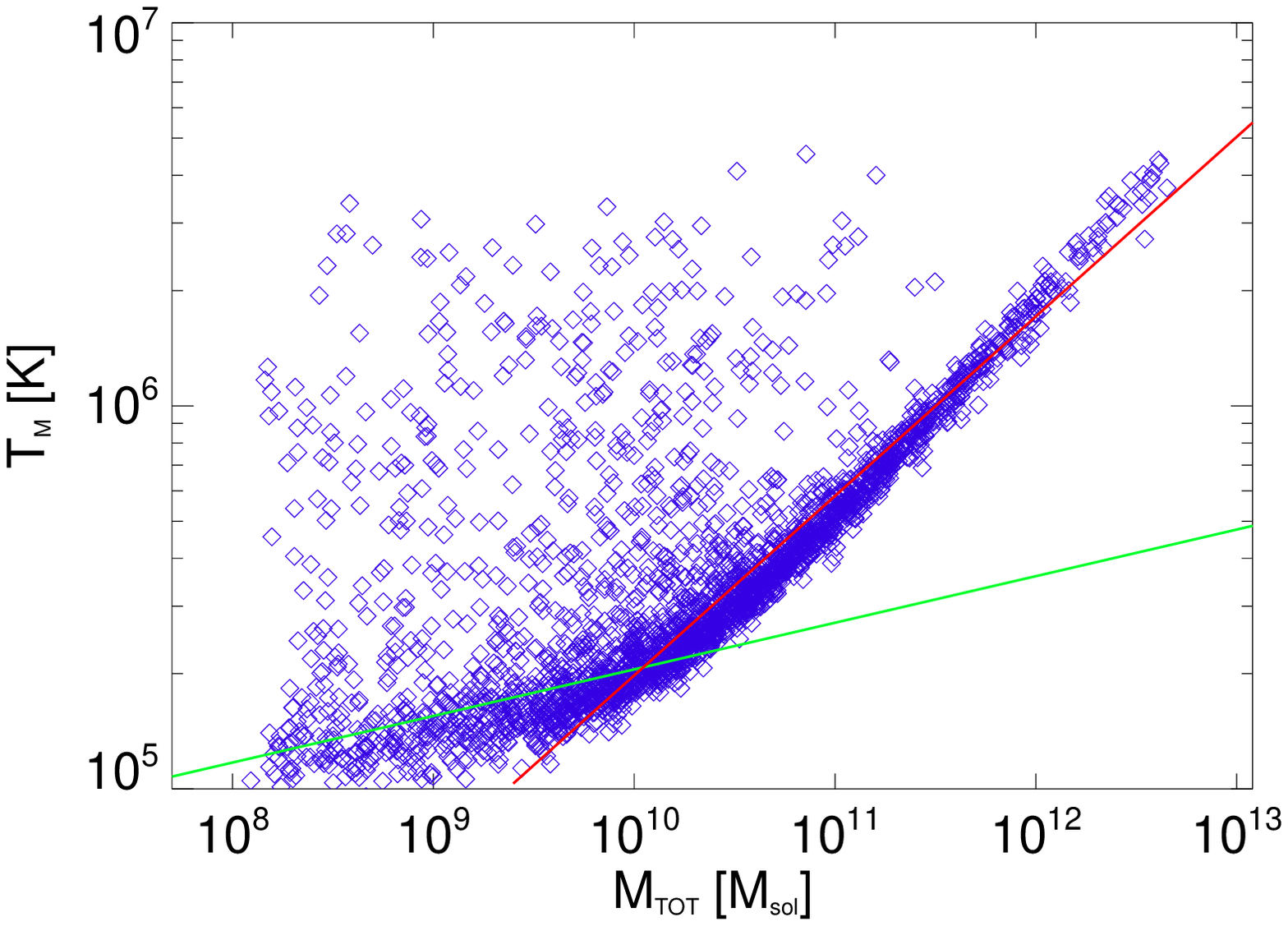}
\end{tabular}
\caption{Radius of simulated galaxies (left panel) and average mass weighted temperature (right panel) 
as a function of their total mass. In the right panel the red and green lines show the best fit calculated for objects
below and above $10^{10}$M$_\odot$ respectively.}
\label{fig:galaxies-th}
\end{figure*}

Fig. \ref{fig:galaxies-th} shows the radius (left panel, spherical symmetry is assumed) and mass-weighted average temperature (right panel) of the simulated galaxies versus their total mass.
The identified objects have masses that range from $10^8$ to more than $10^{12}$M$_\odot$, and 
 radii are between 5 and 200 kpc. 
As it could be expected from the viral theorem, the radius scales almost perfectly with $M^{1/3}$ ($M$ being 
the mass of the galaxy). 
The mass-temperature relation is presented in the right panel of Fig. \ref{fig:galaxies-th}, showing 
three different regimes. For masses bigger than $10^{10}$M$_\odot$, the mass-temperature relation
follows a tight power law with exponent equal to 0.469$\pm$0.008, close to that 
of the whole filament. The temperature ranges between $2-3\times 10^{5}$ and $5\times 10^{6}$K,
significantly higher than those of hosting filaments. 
At smaller temperatures, the relation flattens to an exponent of 0.122$\pm$0.004, 
with temperatures in the range $1-2\times 10^{5}$K, slightly
decreasing toward lower masses. A third region is characterised by a larger scatter in the distribution of temperatures, independent of the
mass of the host galaxies. 
A better understanding of the behaviours of the temperature distribution is given by the following analysis.

\begin{figure*}
\centering
\begin{tabular}{c|c}
  \includegraphics[width=0.45\textwidth]{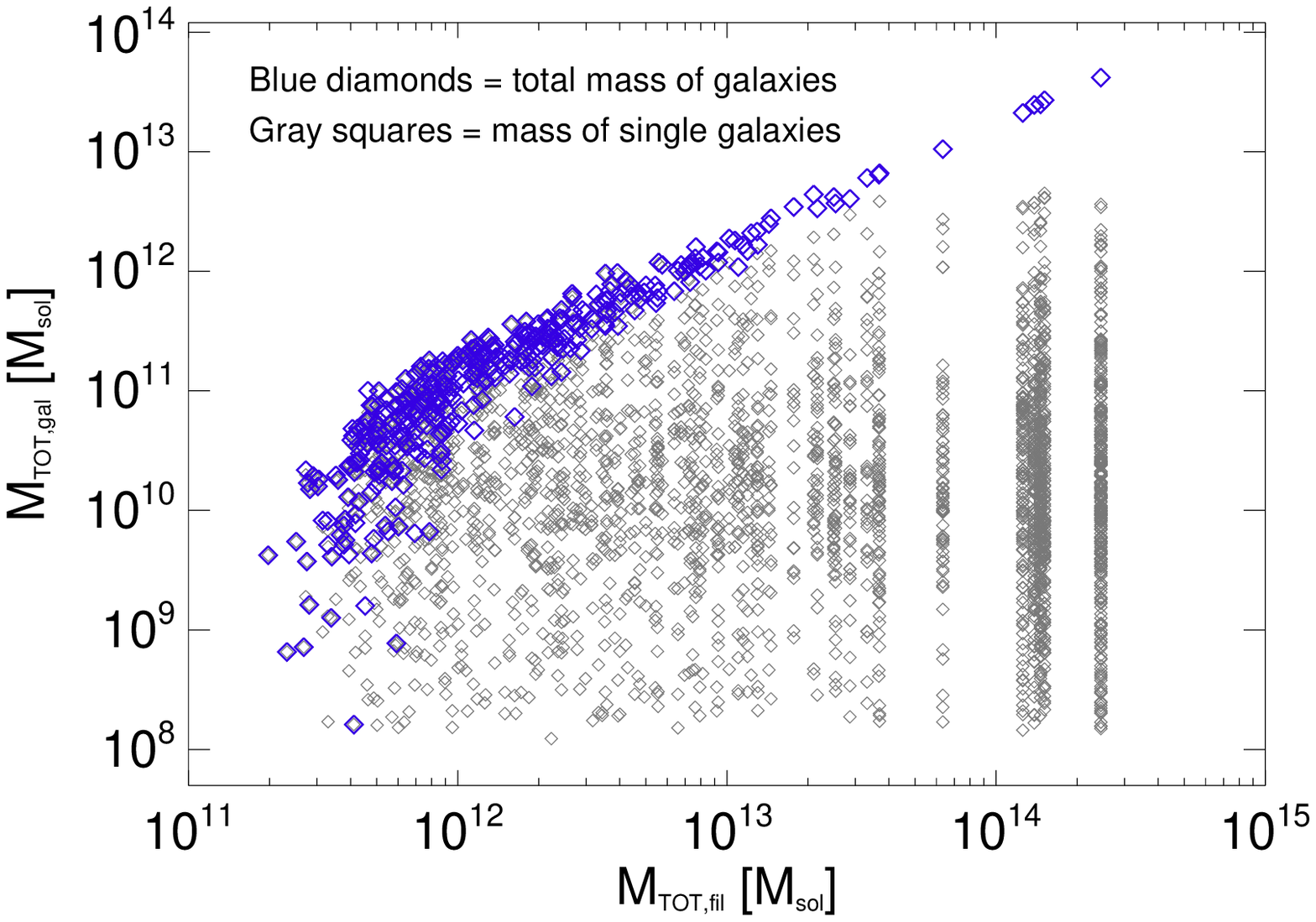} &
  \includegraphics[width=0.45\textwidth]{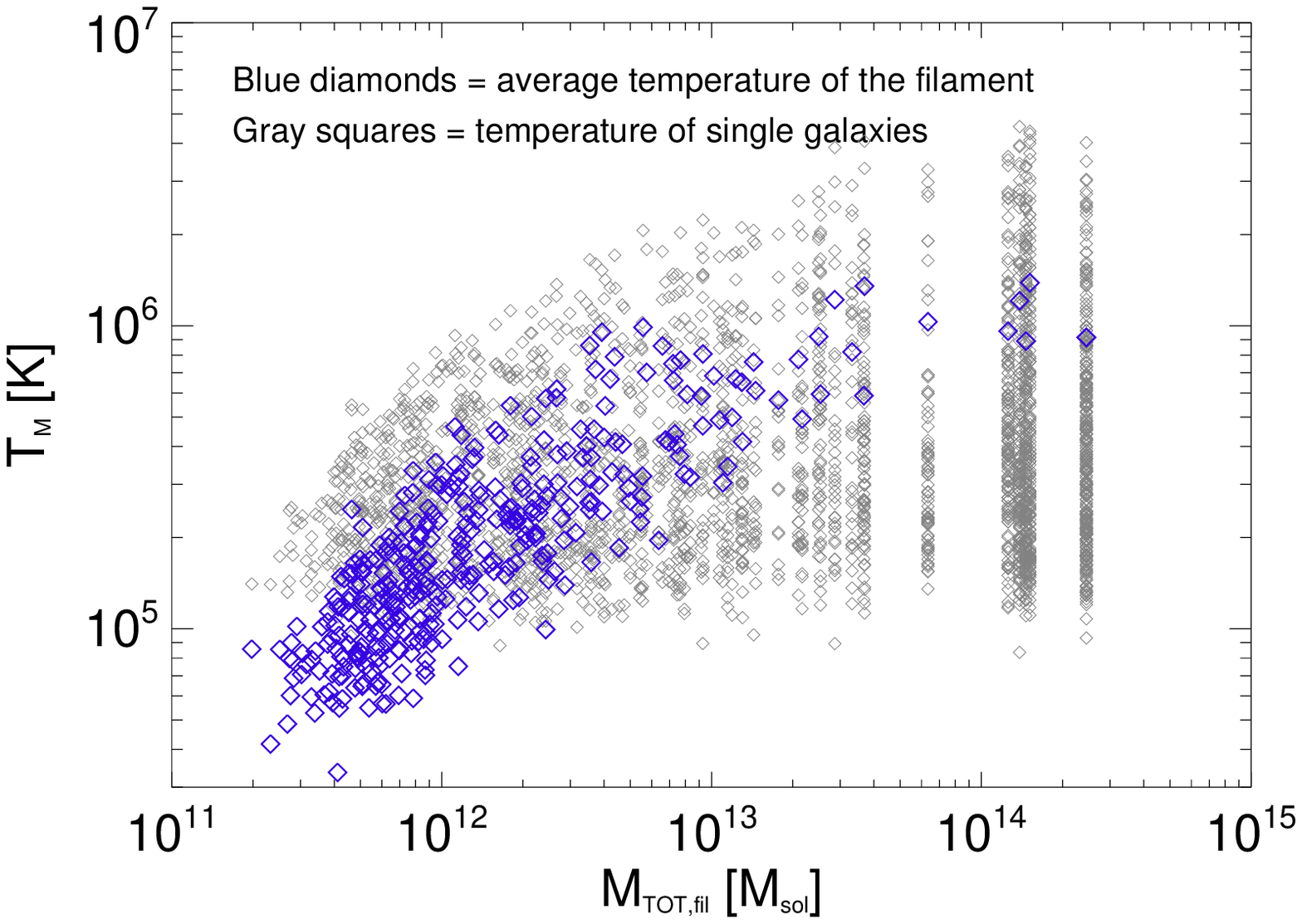}
\end{tabular}
\caption{Total mass (left panel) and average mass weighted temperature (right panel) of the simulated galaxies
as a function of the total mass of the hosting filament.}
\label{fig:galfil}
\end{figure*}

Fig. \ref{fig:galfil} shows how the total mass of galaxies and average mass-weighted temperature
relate to the total mass of the host filament. Two distinctive features emerge. First, 
the maximum galaxy mass and temperature increase with the mass of the filament: the most massive 
and hottest galaxies reside in most massive filaments. The total mass in galaxies presents a remarkably tight correlation
with the total mass of the host filament, and represent a roughly constant $\sim 10\%$ fraction of the total filament mass at all masses. 
 The range of masses and temperatures
increases with the filament mass. This is more evident for the temperature, smaller filaments forming
galaxies with temperatures between 1 and 5$\times 10^5$K, even when the corresponding mass is of the order 
of $10^{11}$M$_\odot$. This suggests that the thermodynamic properties of the galaxies, especially at the lowest masses, 
are strongly influenced by those of the hosting filament. 
Combined with the previous analysis (Fig.~\ref{fig:galaxies-th}), this suggests  
that galaxies with low masses and high temperatures are the result of a strong interaction between the galactic gas and the
one of the host filament, probably during the infall through the strong accretion shocks at the periphery of filaments. 
In these cases, the gas temperature of the galactic halo is equally set by the combination of its internal
pressure equilibrium and of the shock heating by the surrounding environment. For the largest galaxies ($\geq 10^{12} M_{\odot}$) the internal temperature 
follows the viral relation, showing that these objects are residing close to the spine of the host filament and far from strong accretion shocks.
This picture is confirmed by the analysis of the velocity distribution of these galaxies.

\begin{figure*}
\centering
\includegraphics[width=0.7\textwidth]{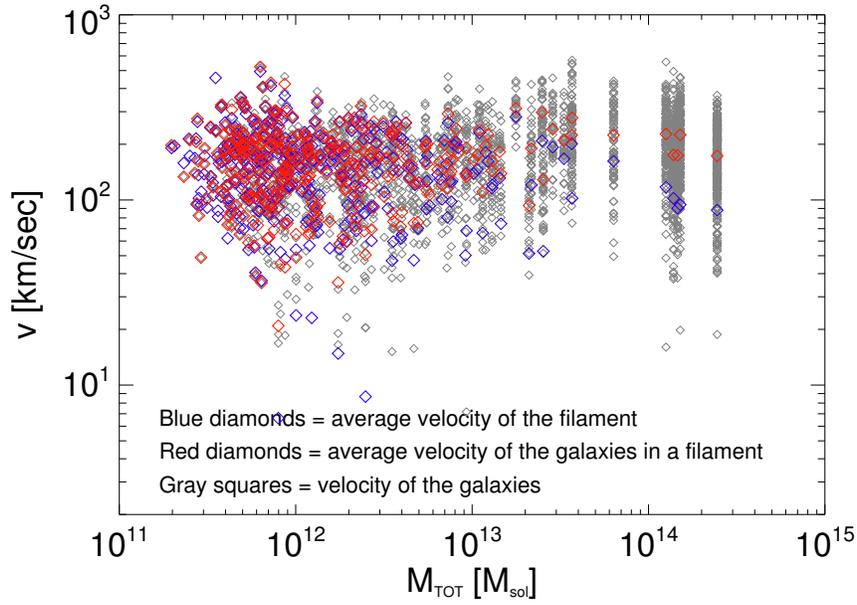} 
\caption{Average velocity of the galaxies' gas compared to that of the hosting filament as a function of 
the host's mass}
\label{fig:velocity}
\end{figure*}

Fig. \ref{fig:velocity} presents the distribution of the average velocity of the gas enclosed in the identified
galaxies as a function of the total mass of the hosting filament. We measure an average velocity of galaxies which is
similar to that of the host filament. We detected a systematic excess by a factor $\sim 2$ of average velocity of the galaxy
population. This is consistent with the fact that accretion shocks are more efficient in thermalising the infall velocity field of the smooth
accreted gas, while the gas close to galaxy halos can keep a significantly larger velocity inside the filament. 
Moreover, a large scatter in the velocity of galaxies is 
observed, and several galaxies have significantly larger relative velocities compared to the surrounding gas. These high velocity objects, which are 
typically found at the periphery of filaments (Fig. \ref{fig:rhoth}-\ref{fig:rhoth2}), are driving shocks while running into the slower
gas within filaments, causing the large temperatures  discussed above.

To conclude, we notice that the relation between galaxies and their host filamentary environment is a very open topic, both in 
observations and simulations. 
Interestingly, \citet{alp15} found that the surrounding environment of 
large-scale structure determines the stellar mass function of galaxies, but otherwise has a modest impact on the galaxy properties.  
On the other hand, using a different algorithm to relate the density distribution of galaxies with large-scale structures in GAMA, \citet{2015MNRAS.448.3665E} reported that  the luminosity function of galaxies is not significantly affected by their large scale structure. 
While our analysis here supports the view that the {\it gas} properties of simulated halos are significantly affected by the properties of their host filament,
our runs lack the necessary small-scale physics to shape the transformation of gas in halos into stars, and therefore must be taken with caution.
In future work we plan to use more complex simulations including radiative cooling, star formation and feedback in order  to investigate the internal properties of simulated galaxies in filaments and to estimate the galaxy-to-gas bias in filaments as a tool to better mark the 
gas distribution of the cosmic web in large surveys of the sky in various wavelengths.

\section{Discussion}
\label{sec:discussion}

\begin{figure*}
\centering
\begin{tabular}{|c|c}
  \includegraphics[width=0.48\textwidth]{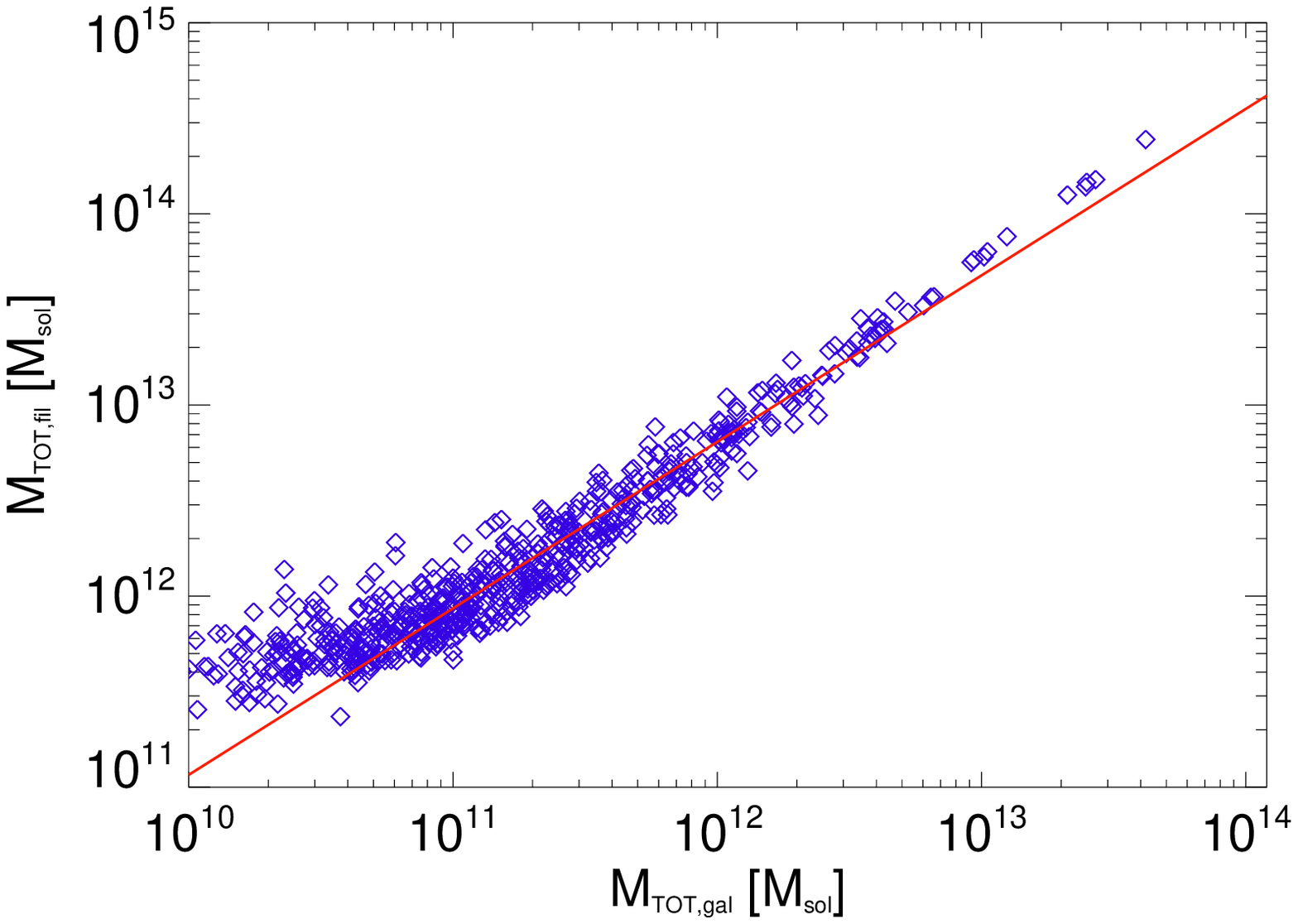} &
  \includegraphics[width=0.48\textwidth]{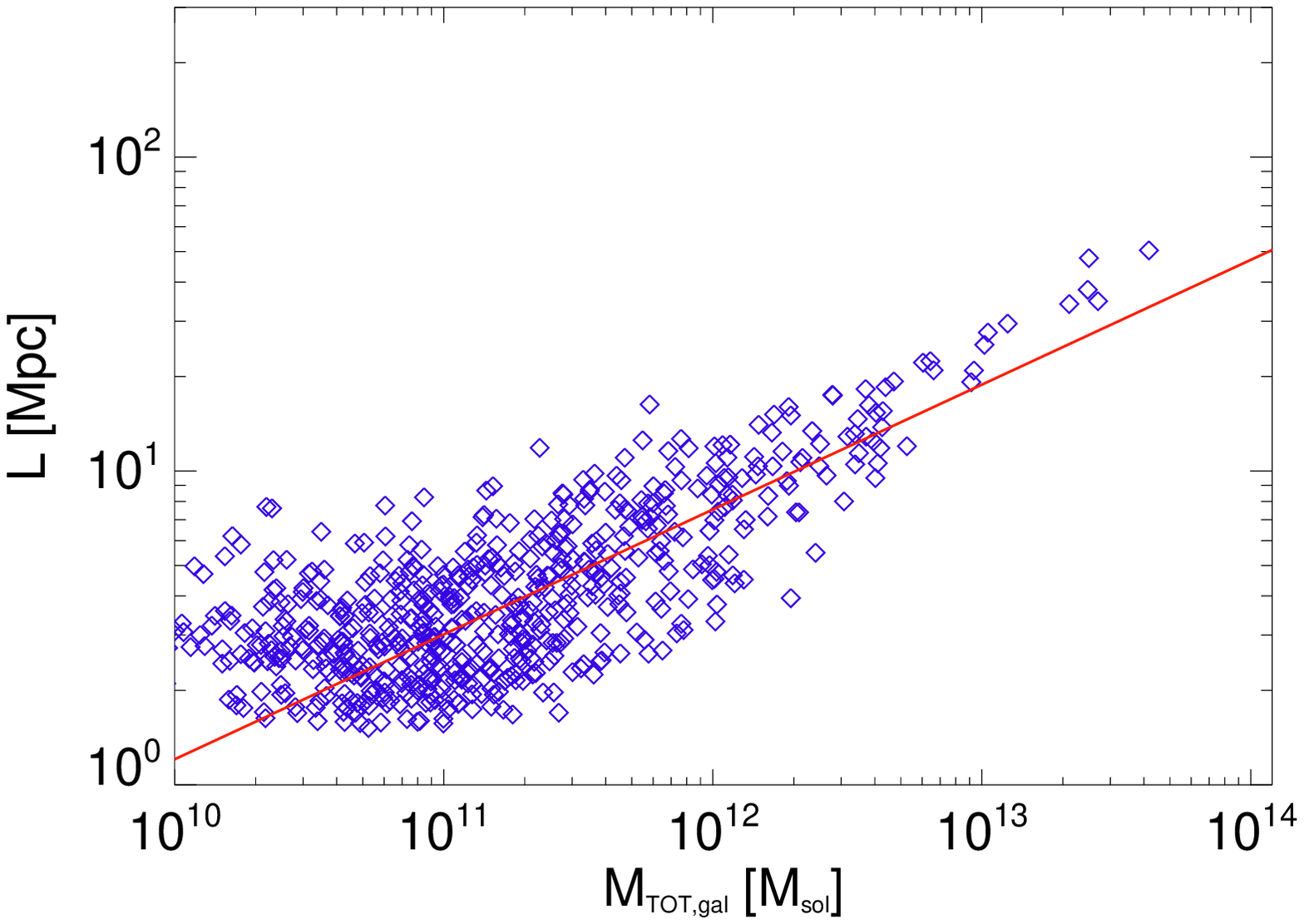} \\
  \includegraphics[width=0.48\textwidth]{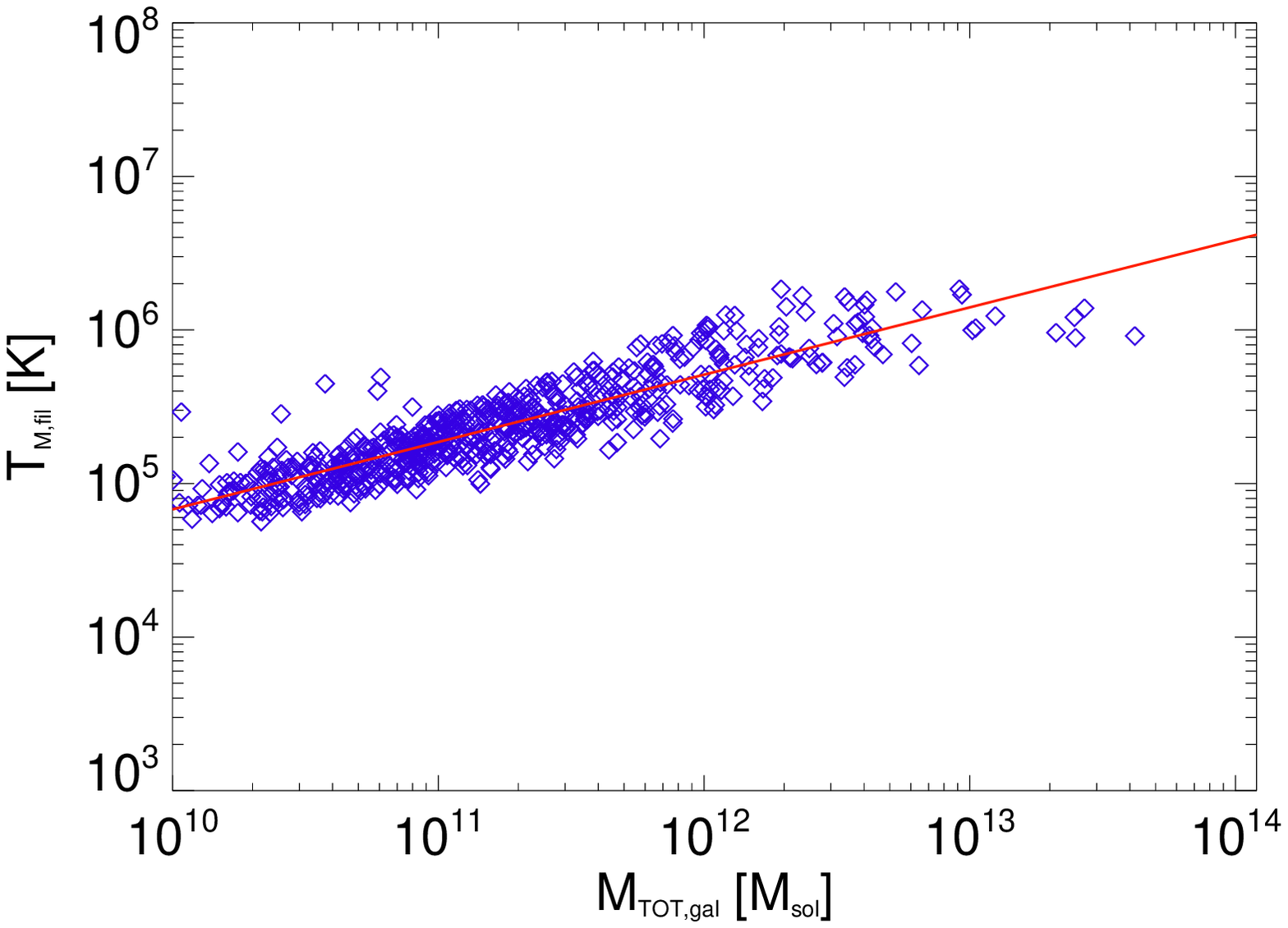} &
  \includegraphics[width=0.48\textwidth]{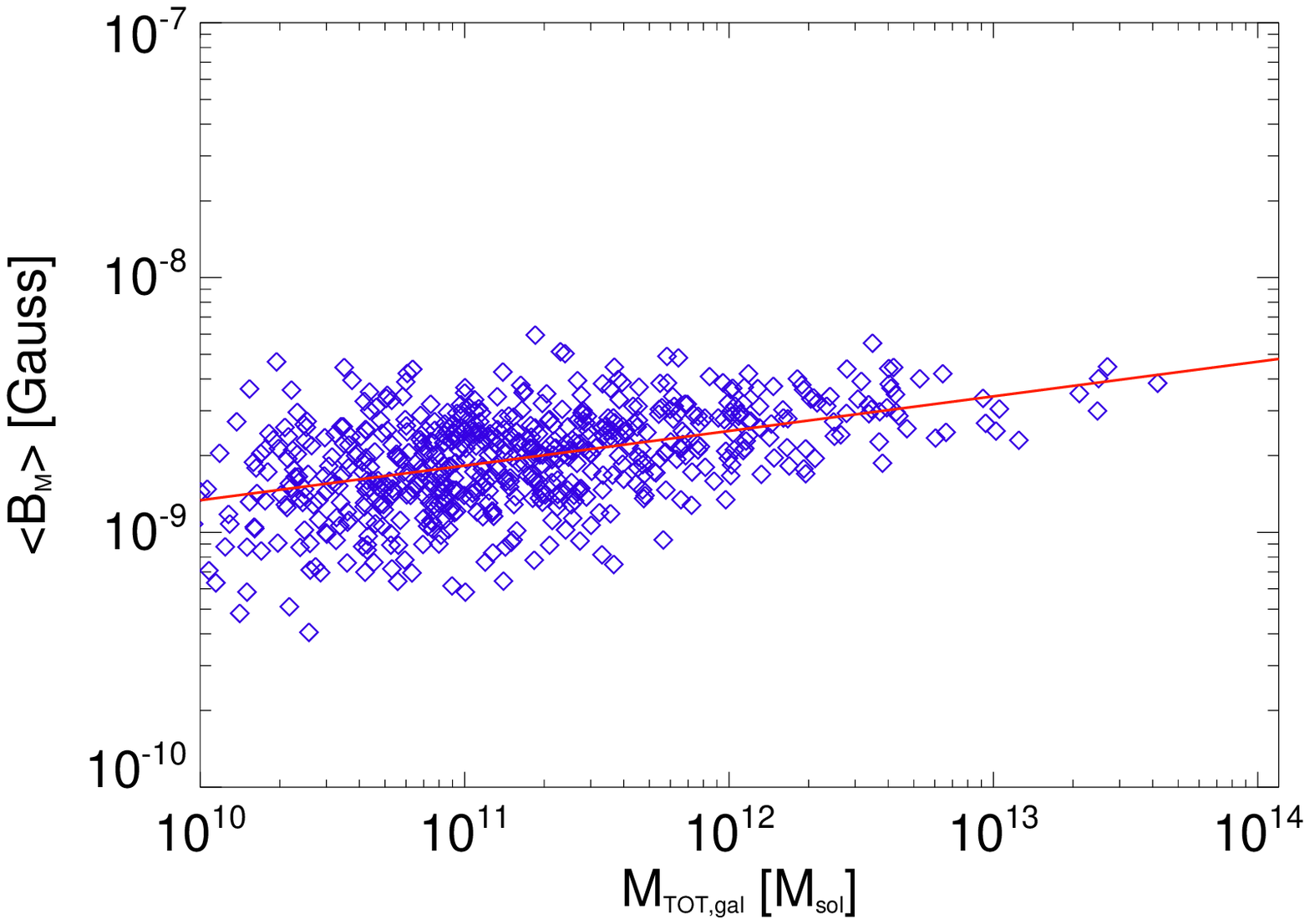}
\end{tabular}
\caption{Correlations between the total mass of simulated galaxies and: a) the total mass of the host filament (top-left); 
b) the length of the host filament (top-right); c) the mass weighted
temperature of the host filaments (bottom left) and d) the mean magnetic field of the host filament (bottom-right). 
Best fit curves are shown as red lines.}
\label{fig:galfit}
\end{figure*}

\begin{table}
\caption{Best-fit parameters of the relations of the total mass of galaxies hosted in a filament and the 
quantities in the first column, that are: the total mass of the hosting filament, its length, its average mass-weighted 
temperature, its average magnetic field intensity. The $\alpha$ parameter is the exponent,
while $\beta$ is the normalization, i.e.  $Y=\beta M_{tot,gal}^{\alpha}$.}
\centering \tabcolsep 5pt
\begin{tabular}{|c|c|c|}
    & $\alpha$ & $\beta$ \\  \hline
   M$_{tot,fil}$  & $0.872 \pm 0.011$ & $2.374 \pm 0.125$\\
   L$_{fil}$       & $0.398 \pm 0.014$ & $-3.896 \pm 0.163$\\
   T$_{avg,fil}$  & $0.438 \pm 0.011$ & $0.445 \pm 0.135$\\
   B$_{avg,fil}$  & $0.136 \pm 0.013$ & $-10.233 \pm 0.150$\\
\end{tabular}
\label{tab:finalfit}
\end{table}

\subsection{Properties of filaments}
\label{subsec:disc1}
Some properties of simulated filaments may depend on numerical effects such as the finite spatial/mass resolution
of our runs. The use of our different volumes provides a mean  to address these effects. 

The most massive and volume-filling filaments are formed in the largest box
(BOX200) with fluctuations on larger scales driving the growth of more massive objects. Numerical diffusion opposes the infall of matter and makes structures more diffuse in lower resolution runs, making filaments slightly more volume-filling compared to more resolved filaments of the same mass (Sec.  \ref{subsec:thermo}).
Spatial resolution also affects the temperature of the filaments, leading to increasingly
higher average temperature objects as the resolution decreases.
Overall, box size and resolution affect the volume/length/temperature vs. mass relations
in terms of normalization of the different distribution, within a factor $\sim 2$, while the slope of the best-fit relations is only marginally affected 
(see Sec.  \ref{subsec:thermo}). 
In particular, the mass-temperature relation is  compatible with
what is expected for virialized cylindrical objects.

The baryon fraction in most filaments is biased low at low resolution, while we find that at all resolutions objects
with  $10^{13} \rm M_{\odot}$ are instead more independent of resolution in our runs, and shows no evolution for $z \leq 1$ (see Sec.  \ref{subsec:thermo}).
The magnetic field in filaments shows the largest dependence on resolution effects (see Sec.  \ref{subsec:magnetic}). Our previous MHD results in \citet{va14mhd} show that the amplification of magnetic fields in filament very slowly increases
 with resolution, even if no small-scale dynamo is produced. At increased resolutions, the average magnetic field in filaments shows a slow increase with mass. On the other hand, the maximum magnetic field within filaments is more significantly increased with mass and resolution, consistent with the picture that turbulent flows are better
 captured by simulations in larger objects and at higher resolution. The predominance of supersonic turbulent forcing (see Sec. \ref{subsec:profiles}) makes small-scale dynamo in filaments very unlikely to occur \citep[e.g.][]{ry08,va14mhd}.
 
We have presented how different statistical relations depend on time, analysing the properties of filaments  at
three different redshift, z=0.0, z=0.5, z=1.0.  We report only 
a mild time evolution at z$\le$1. The main changes with time are essentially due to the
gravitational infall process, which leads to more compact structures. An important role is played by the 
 progressive
formation of galaxy clusters at z$\leq$1, which affects our identification of filaments.  The formation of a galaxy cluster at low redshift can break a larger
network of connected structures, which are then identified as  individual entities.

\subsection{Properties of galaxies in filaments}

Our simulated population of galaxies in filaments shows a remarkable similarity with the galaxies in filaments identified in the GAMA survey \citep[][]{alp14a}. 
Simulated filaments longer than $\sim 10$ Mpc host the same number of galaxies as observed ones. For a given filament mass the total and the average stellar
mass in galaxies is consistent with observations. Also other properties such as masses and radii are compatible with galaxies (or small groups of galaxies).\\

Although filaments contain a broad variety of galaxies in terms of mass, average temperature and
velocity, there is a strong dependence of these properties on the
characteristics of the hosting filament. 
The total mass of galaxies in a filament is nearly a constant fraction ($\sim 10\%$) of the mass of the host filament, and while the internal gas temperature of simulated galaxies mostly depends on the mass of the 
host galaxy, in a significant fraction of low mass galaxies their internal temperature is linked to that of the host filament via shock heating.  This is found in small halos characterised
by large infall velocities onto the host filament, while larger objects are typically found within accretion shocks and move along the spine of filaments (Fig. \ref{fig:rhoth}-\ref{fig:rhoth2}).
An important caveat of this analysis is that several assumptions on the star-to-DM mass ratio are necessary both in observations and simulations (Sec. \ref{sec:galaxy}). 
From the numerical point of view, several important ingredients are missing from our modelling of galaxies, and future work will be necessary to assess their impact on the 
distribution of internal properties of galaxies.
Increased spatial and forces resolution might change the stripping and the shock heating of galaxies forming in filaments, affecting their thermodynamical properties \citep[e.g.][]{2007MNRAS.380.1399R,2009A&A...499...87K} as well as by quenching their star formation \citep[e.g.][]{2016MNRAS.455..127M}.
On the other hand, radiative gas cooling and radiative/momentum feedback from the galactic activity should affect the internal temperature of galaxies, wiping the contribution from shocks in filaments
at large over densities \citep[e.g.][]{2015MNRAS.454...83W,2016MNRAS.458..270R}.  Moreover, galactic feedback should also contribute to the magnetisation of the WHIM within some distance from galaxies \citep[e.g.][]{donn09,2016MNRAS.456L..69M}.

\subsection{Observational implications}
\label{subsec:observations}

The tight correlation between the galaxy population in filaments and the host filament allows derivation of other correlations between the galaxy properties and the properties of the
surrounding gas. This is  a crucial step that can predict the thermal and magnetic properties of the WHIM on $\sim 10-100$ Mpc scales, based on optical/IR survery targeting the stellar mass component of galaxies. 
In the panels of Fig.  \ref{fig:galfit} we derive for BOX50 the correlations between the total mass of simulated galaxies and: a) the total mass of the host filament; b) the length of the host filament; c) the mass weighted
temperature of the host filaments and d) the mean magnetic field of the host filament. The best fit of the $\log(Y)=A \log(M_{\rm tot,gal}) + B$ for the different relations are given in Tab. \ref{tab:finalfit}.
For objects with a total mass in galaxies larger than $10^{11} M_{\odot}$ all correlations are narrow and characterised by a constant scatter smaller than $\sim 0.5$ dex. 
Based on the trends with resolutions (Sec. \ref{subsec:disc1}), the same correlations should also be extended to larger masses, not contained 
in the BOX50 run. 
Given the very good match between the simulated and the observed galaxies in the GLSSC catalogue of the GAMA survey (Sec. \ref{sec:galaxy}), we can now surmise that these 
correlations can be used to predict the basic emission properties of the gas in filaments. For a filament of a given mass our profiles can give the
probability of detecting a certain X-ray emission level from a fraction of its volume, and relate the emission to the underlying WHIM mass  \citep[][]{2015A&A...583A.142N}.  Similarly, the knowledge of the average magnetic field in filaments as a function of the observed properties of galaxies
can enable a better prediction of the expected synchrotron emission from the shocked WHIM, to be observed by future radio surveys or with long exposures \citep[e.g.][]{va15ska,va15radio}.\\

These correlations can be naively interpreted as the result of the obvious fact that larger objects form more galaxies, and that these galaxies form on the high density peaks of the total matter distribution
that surrounds them, which in turn set the thermal properties of the diffuse gas via shock heating. Nevertheless,  the fact that the net outcome of the overall star formation within galaxies  (a process that ultimately operates on $\leq $pc scales) correlates so tightly with the average gas temperature on structures up to $\sim 10^2$ Mpc long, highlights the role of self-gravity in organising the properties of cosmic objects across a tremendous range of scales ($\geq 10^8$).

\section{Conclusions}
\label{sec:conclusions}

Cosmological filaments permeate the universe and account for a meaningful fraction of its mass. 
However, their detection is extremely challenging, due to their physical properties, which make
them almost invisible at any wavelength. Therefore, it is crucial to find reliable approaches
to identify a filament and characterize its main properties through easily detectable tracers. This would not only
facilitate the filaments discovery process, but would also give essential hints to 
tune and optimize instruments and observations finalized to their detection.

Our conclusions can be summarised as follows:

\begin{itemize}
\item In the redshift range explored ($0 \leq z \leq 1$) the thermal and magnetic properties evolve only mildly, e.g. the average temperature of the most massive objects decreases by a factor $\sim 2$ from z=0 to z=1 and also the mean and maximum magnetic fields evolve similarly. The length of filaments show a slow increase in this redshift range.
This suggests that the properties of cosmic filaments are set already by $z \approx 1$, i.e. at a look-back time of $\sim 8$ Gyr. This is consistent with the fact that they are objects characterised by a very early formation epoch, and that their dynamical evolution is very slow. 

\item We find a small impact of the spatial/mass resolution on the global properties of the simulated population of filaments, with a factor $\sim 2$ in most cases. As resolution is increased, the thermalisation of filaments tends to decrease (due to the weakening of accretion shocks), the average length at a given mass is increased for the formation of more low density branches around filaments, and the magnetic field at a given mass is increased by the enhanced compression within filaments. 

\item  The volume and density profiles of filaments show that the most important quantities (temperature, velocity, X-ray luminosity, magnetic fields) show a remarkable self-similar behaviour across the explored mass range. 

\item Galaxies provide an effective proxy to characterize the geometric and physical properties 
of the hosting filaments, facilitating their detection through future observational campaigns. Future simulations including a more detailed description of galactic physics and higher spatial resolution
will give even more accurate and comprehensive indications. 
\end{itemize}

\section*{acknowledgements}

Computations accomplished in this work were performed using the {\enzo} code (http://enzo-project.org), 
which is the product of a collaborative effort of scientists at many universities and national laboratories. 
We gratefully acknowledge the {\enzo} development group for providing helpful and well-maintained on-line documentation and tutorials.
We also acknowledge ETHZ-CSCS{\footnote{http://www.cscs.ch}} for the use of the Piz Daint systems in the Chronos project ID ch2 and s585,
and of the Piz Dora system for data processing and visualization.
We would like to thank Jean Favre for the support on VisIt and Maria Grazia Giuffreda for her valuable technical assistance at CSCS.
FV acknowledges personal support
from the grant VA 876/3-1 from the Deutsche Forschungsgemeinschaft (DFG).
FV and MB acknowledge partial support from
the grant FOR1254 from DFG. We also acknowledge the use of computing resources under allocations
no. 7006 and 9016 (FV) and 9059 (MB) on supercomputers at the
NIC of the Forschungszentrum J\"{u}lich.
MA is funded by an appointment to the NASA Postdoctoral Program at Ames Research Centre, administered by Universities Space Research Association through a contract with NASA.
GAMA is a joint European-Australasian project based around a spectroscopic campaign using the Anglo-Australian Telescope. The GAMA input catalogue is based on data taken from the Sloan Digital Sky Survey and the UKIRT Infrared Deep Sky Survey. Complementary imaging of the GAMA regions is being obtained by a number of independent survey programs including GALEX MIS, VST KiDS, VISTA VIKING, WISE, Herschel-ATLAS, GMRT and ASKAP providing UV to radio coverage. GAMA is funded by the STFC (UK), the ARC (Australia), the AAO, and the participating institutions. The GAMA website is http://www.gama-survey.org/. The VISTA VIKING data used in this paper is based on observations made with ESO Telescopes at the La Silla Paranal Observatory under programme ID 179.A-2004.

\bigskip

\bibliographystyle{mnras}
\bibliography{filaments2}

\end{document}